\documentclass[sigconf]{acmart}

\usepackage{booktabs} % For formal tables

\usepackage{balance}

\usepackage{mdwlist}% Compact list formats \itemize*, \enumerate*

\renewcommand\footnotetextcopyrightpermission[1]{}
\setcopyright{none}
\acmDOI{.}
\acmISBN{.}
\acmConference[ESEC/FSE]{ESEC/FSE 2017}{}{}
\acmYear{2017}
\copyrightyear{2017}
\acmPrice{.}

\usepackage{algorithm}
\usepackage[noend]{algpseudocode}    % for algorithmic
\usepackage{bussproofs}
\algrenewcommand\algorithmicindent{0.7em}%

\usepackage{microtype}
\usepackage{adjustbox}
\usepackage{flushend}
\usepackage[USenglish]{babel}
\usepackage{paralist}
\usepackage{amsmath}
\usepackage{enumitem}
\usepackage{amsfonts}
\usepackage{amsthm}
\usepackage{bussproofs}
\usepackage{varwidth}
\usepackage{listings}
\usepackage{booktabs}
\usepackage{amssymb}

\newcommand{\ignore}[1]{}

\newcommand{\textcode}[1]{\texttt{#1}}
\newcommand{\toolname}{\textsc{FruitTree}}
\newcommand{\Watts}{\textsc{Watts}}
\newcommand{\Duet}{\textsc{Duet}}
\newcommand{\US}{$^\bigstar$}

\newcommand{\tfunc}{\textsf{TFunc}}

\newcommand{\AnalyzeTM}{\textsf{AnalyzeTM}}

\newcommand{\AnalyzeSeq}{\textsf{AnalyzeSeq}}
\newcommand{\Interfs}{\textsf{Interfs}}

\newcommand{\Analyze}{\textsf{Analyze}}
\newcommand{\AnalyzeAll}{\textsf{AnalyzeAll}}

\newcommand{\FilterFeasible}{\textsf{FilterFeasible}}

\newcommand{\map}{\textsf{map}}

\newcommand{\lfp}{\textsf{lfp}}
\newcommand{\JoinMM}{\textsf{JoinMM}}

\newcommand{\GraphSet}{\mathbb{G}}
\newcommand{\NodeSet}{\mathbb{N}}
\newcommand{\StateSet}{\mathbb{S}}
\newcommand{\VarSet}{\mathbb{V}}

\newcommand{\GraphNodes}{\textsf{n}}
\newcommand{\GraphTrans}{\textsf{t}}

\newenvironment{mathprooftree}
  {\varwidth{.9\textwidth}\centering\leavevmode}
  {\DisplayProof\endvarwidth}
\usepackage{stackengine}

\newcommand\mknote[1]{\textcolor{red}{{\textbf{Markus Says: #1}}}}

\begin{document}

\title{Thread-Modular Static Analysis for Relaxed Memory Models}

\author{Markus Kusano}
\affiliation{%
  \institution{Virginia Tech}
  \city{Blacksburg} 
  \state{VA,} 
  \postcode{USA}
}
%\email{mukusano@vt.edu}

\author{Chao Wang}
\affiliation{%
  \institution{University of Southern California}
  \city{Los Angeles} 
  \state{CA,} 
  \postcode{USA}
}
%\email{wang626@usc.edu}

\begin{abstract}

We propose a memory-model-aware static program analysis method for
accurately analyzing the behavior of concurrent software running on
processors with weak consistency models such as x86-TSO, SPARC-PSO,
and SPARC-RMO.  At the center of our method is a unified framework for
deciding the feasibility of inter-thread interferences to avoid
propagating spurious data flows during static analysis and thus boost
the performance of the static analyzer.  We formulate the checking of
interference feasibility as a set of Datalog rules which are both
efficiently solvable and general enough to capture a range of
hardware-level memory models.  Compared to existing techniques, our
method can significantly reduce the number of bogus alarms as well as
unsound proofs.  We implemented the method and evaluated it on a large
set of multithreaded C programs.  Our experiments show the method
significantly outperforms state-of-the-art techniques in terms of
accuracy with only moderate runtime overhead.

\end{abstract}

%
% The code below should be generated by the tool at
% http://dl.acm.org/ccs.cfm
% Please copy and paste the code instead of the example below. 
%

%
% End generated code
%

%
%  Use this command to print the description
%
%\printccsdesc

% We no longer use \terms command
%\terms{Theory}


%ommit page numbering.
\pagenumbering{gobble}

\maketitle

\section{Introduction}

Concurrent software written for modern computer architectures, though
ubiquitous, remains challenging for static program analysis. Although
abstract interpretation~\cite{Cousot77} is a powerful static analysis
technique and prior \emph{thread-modular}
methods~\cite{Ferrara08,Mine11,Mine12,Mine14,KusanoW16} mitigated
\emph{interleaving explosion}, none was specifically designed for
software running on weakly consistent memory.
%
%This is a serious deficiency since processors with weakly consistent memory may
%exhibit behaviors not permitted by uniprocessors.  
This is a serious deficiency since weakly consistent memory may exhibit
behaviors not permitted by uniprocessors.
%
%For example, it may delay or reorder slow memory accesses to increase
%performance, thereby introducing additional inter-thread non-determinism.  
For example, slow memory accesses may be delayed, increasing
performance, but also introducing additional inter-thread non-determinism.
Thus, multithreaded software running on such 
processors may exhibit erroneous behaviors not manifesting on sequentially
consistent (SC) memory.

Consider x86-TSO (total store order) as an example.  Under TSO, each processor
has a \emph{store buffer} caching memory write operations so they do not block
the execution of subsequent instructions~\cite{AdveG96}.  
Conceptually, each processor has a queue of pending writes to be flushed
to memory at a later time. 
The flush occurs non-deterministically at any time during the program's
execution.
This delay between the time a write instruction executes and the time it takes
effect may cause the write to appear reordered with subsequent instructions
within the same thread.
Figure~\ref{fig:dekker} shows an example where the assertion holds
under SC but not TSO.  Since $x$ and $y$ are initialized to 0 and they
are \emph{not} defined as \emph{atomic} variables, the write
operations (\texttt{x=1} and \texttt{y=1}) may be stored in buffers,
one for each thread, and thus delayed after the read operations.

SPARC-PSO (partial store order) permits even more non-SC behaviors: it
uses a separate store buffer for each memory address.
That is, \texttt{x=1} and \texttt{y=1} within the same thread may be
cached in different store buffers and flushed to memory independently.
This permits the reordering of a write to $x$ with a
subsequent read from $y$, but also with a subsequent write
(e.g., to variable $z$) in the same thread.
The situation is similar under SPARC-RMO (relaxed-memory order).  
We detail how such relaxation leads to errors in Section~\ref{sec:motivation}.
%We will show in Section~\ref{sec:motivation} that such
%behavior may lead to erroneous program behaviors.

\begin{figure}
\vspace{1ex}
\centering

{\footnotesize

\begin{minipage}{0.425\linewidth}
\begin{lstlisting}[language=C,frame=single]
void thread1() {
  x = 1;
  a = y;
}
\end{lstlisting}
\end{minipage}
\hspace{0.10\linewidth}
\begin{minipage}{0.425\linewidth}
\begin{lstlisting}[language=C,frame=single]
void thread2() {
  y = 1;
  b = x;
}
\end{lstlisting}
\end{minipage}

\textcode{assert( !(a == 0 \&\& b == 0) );}
}

\vspace{-1ex}
\caption{The assertion holds under SC, but not under x86-TSO, SPARC-PSO, and SPARC-RMO memory models.}
\label{fig:dekker}
\vspace{-4ex}
\end{figure}

%Broadly speaking, existing \emph{thread-modular} methods for abstract
%interpretation of concurrent programs fall into two categories, none
%of which models weak-memory related behaviors.  
%
Broadly speaking, existing \emph{thread-modular} abstract interpreters fall
into two categories, neither modeling weak-memory related behaviors.
The first are {SC-specific}~\cite{KusanoW16,FarzanK12,FarzanK13}: they are
designed to be flow-sensitive in terms of modeling thread interactions but
consider only behaviors compatible with the SC memory.  
The second~\cite{Mine11,Mine12,Mine14} are oblivious to memory models
(MM-oblivious): they permit all orderings of memory-writes across threads.
Therefore,
MM-oblivious methods may report spurious errors (bogus alarms) whereas
SC-specific methods, although more accurate for SC memory, may miss real
errors on weaker memory (bogus proofs).  
This flaw is not easy to fix using conventional approaches~\cite{Mine14}. For
example, maintaining relational invariants at all program points makes the
analysis prohibitively expensive.
In Section~\ref{sec:motivation}, we use examples to illustrate issues
related to these techniques.

\begin{figure}
\vspace{-2ex}
\centering
\includegraphics[width=.9\linewidth]{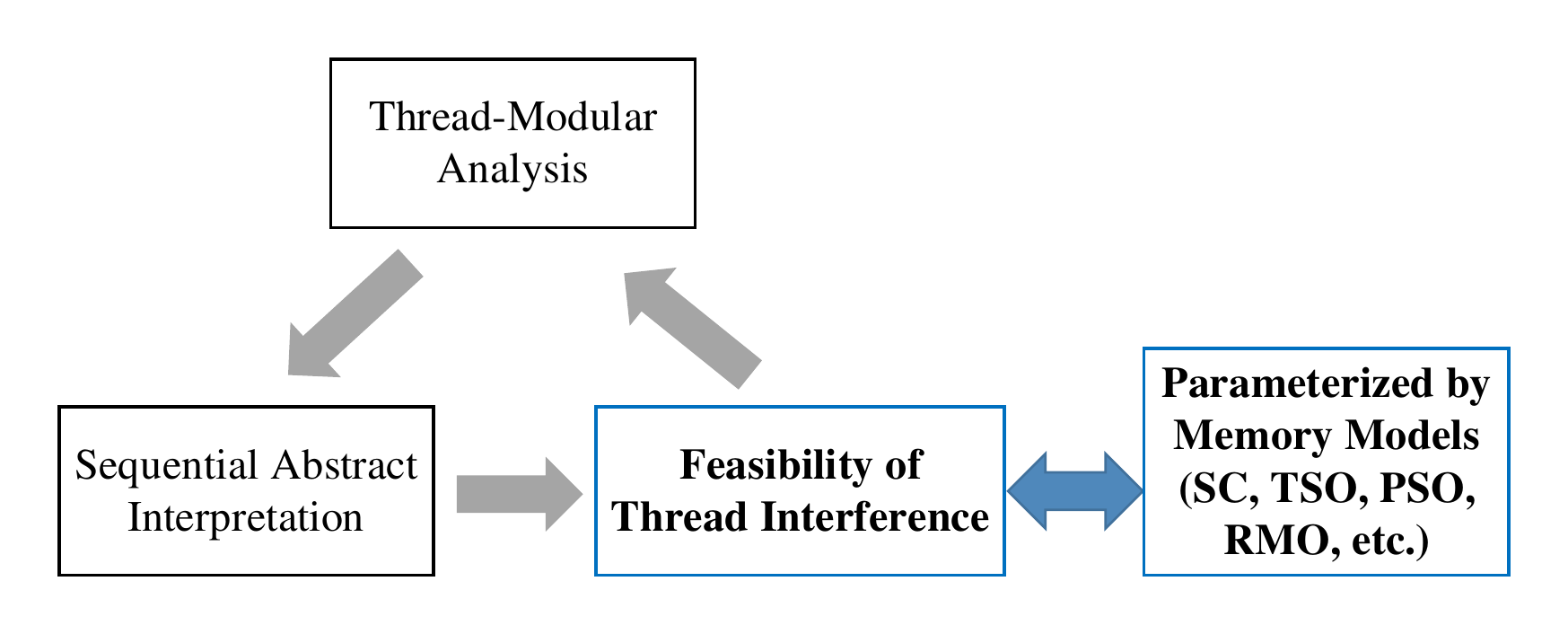}
\vspace{-2ex}
\caption{\toolname{}:  Our memory-model-aware, thread-modular, static program analysis procedure.}
\label{fig:diagram}
\vspace{-3ex}
\end{figure}

We propose the first thread-modular abstract interpreter for analyzing
concurrent programs under weakly consistent memory.  
%
%Our method is both {flow-sensitive} in modeling thread interactions and
%\emph{memory-model specific}: its characterization of the memory operations is
%specific to each given processor-level memory model of interest, as shown in
%Figure~\ref{fig:diagram}.  
%
Our method models thread interactions with flow-sensitivity, and is
memory-model specific: it models memory operations assuming a
processor-level memory model, as shown in Figure~\ref{fig:diagram}.
In this figure, the boxes with bold text highlight our main contributions.

%As shown in Figure~\ref{fig:limitation}, when applied to the TSO
%memory model, our method has the advantage of being more accurate than
%existing methods: it avoids the bogus behavior (incorrectly) allowed
%by MM-oblivious methods~\cite{Mine11,Mine12,Mine14} and thus
%eliminates false positives. It also captures the real behavior
%incorrectly omitted by SC-specific methods~\cite{KusanoW16} and thus
%eliminates falsely verified proprieties.
%
%
%\begin{figure}
%\centering
%\includegraphics[width=0.9\linewidth]{methods}
%\caption{Limitations of existing abstract interpretation-based static
%  analysis techniques for relaxed memory models such as TSO.}
%\label{fig:limitation}
%\end{figure}

%One advantage of our method is that it builds on a new and unified
%framework for modeling consistency semantics. 
%
Our method builds on a unified framework for modeling
the memory consistency semantics.
%
%Within this framework, deciding the feasibility of thread interactions is
%formulated as a constraint problem (via Datalog), which is efficiently solvable
%and can determine if the interactions are allowed by each memory model.
%
Specifically, the feasibility of thread interactions is formulated as a
constraint problem via Datalog: it is efficiently solvable in polynomial time,
and adaptable to various hardware-level memory models.
Additionally, our method handles thread interactions in a
flow-sensitive fashion while being thread-modular.  
Analyzing one thread at a time, as opposed to the entire program, increases
efficiency, especially for large programs.
However, unlike prior MM-oblivious methods  we do not join all the effects of
remote stores before propagating them to a thread, thus preserving
accuracy.
Overall, our method differs from the state-of-the-art,
which either are \emph{non-thread-modular}~\cite{KupersteinVY11,FarzanK12,Meshman2014,Dan15}
or not specifically targeting weak
memory~\cite{Mine11,Mine12,Mine14,KusanoW16}.

Our method also differs significantly from techniques designed for bug
hunting as opposed to obtaining correctness proofs.
For example, in concurrency testing, stateless dynamic model
checking~\cite{Godefr97,Flanag05} was extended from SC to weaker
memory
models~\cite{NorrisD13,ZhangKW15,AbdullaAAJLS15,DemskyL15,Huang16a,AbdullaAJL16,OuD17}.
In bounded model checking, Alglave et al.~\cite{Alglave13T} modeled
weak memory through code transformation or direct symbolic
encoding~\cite{Alglave13P,Alglave14}.
However, these methods cannot be used to verify properties: if they do
not find bugs, it does not mean the program is correct.
In contrast, our method, like other abstract interpreters, is geared toward
obtaining correctness proofs. 
%
%Thus, it is complementary to these existing techniques.

We implemented our new method in a tool named \toolname{}, using
Clang/LLVM~\cite{AdveLBSG03} as the C front-end,
Apron~\cite{Jeannet09} for abstract domains, and the
$\mu$Z~\cite{Hoder11} Datalog engine in Z3~\cite{DeMoura08}.
We evaluated \toolname{} on 209 litmus tests, and 52 larger multithreaded
programs totaling of 61,981 lines of C code. Reachability properties were
expressed as embedded assertions.
Our results show that \toolname{} is significantly more accurate than
state-of-the-art techniques with moderate runtime overhead.

Specifically, we compared \toolname{} against the MM-oblivious
analyzer of Min\'{e}~\cite{Mine14}, the SC-specific thread-modular
analyzer \Watts{}~\cite{KusanoW16}, and a non thread-modular analyzer
named \Duet{}~\cite{FarzanK12,FarzanK13}.
On the litmus tests, \toolname{} is more accurate than the other three
methods.  
On the larger benchmarks, including Linux device drivers, \toolname{}
proved 4,577 properties, compared to 1,752 proved by Min\'e's method.
%
% markus: should we even mention the number of properties proved by Duet and
% Watts since they are unsound? If we are talking about verification then if
% the tool is unsound then it proves zero properties. This may reduce
% confusion.
%\toolname{} even proved more properties than the potentially-unsound
%tool \Duet{} and came close to the same amount as the unsound tool
%\Watts{} (unsound for weak memory), which proved 4,853 and 2,432
%properties, respectively.

To summarize, we make the following contributions:
\begin{itemize*}
\item 
We propose a memory-model aware static analysis method
based on thread-modular abstract interpretation.
\item 
We introduce a declarative analysis framework for deducing the
feasibility of thread-interferences on weak memory.
\item 
We implement and evaluate our method on a set of benchmarks to
demonstrate its high accuracy and moderate runtime overhead.
\end{itemize*}

The remainder of this paper is organized as follows. 
First, we motivate our technique via examples in Section~\ref{sec:motivation}.
Then, we provide background on memory models and abstract interpretation in
Section~\ref{sec:background}.
We present our new declarative analysis for checking the feasibility
of thread inferences in Section~\ref{sec:feasibility}, followed by the
main algorithm for thread-modular abstract interpretation in
Section~\ref{sec:algorithm}.
We present our experimental results in Section~\ref{sec:experiment},
review related work in Section~\ref{sec:related}, and conclude in
Section~\ref{sec:conclusion}.

\section{Motivation}
\label{sec:motivation}

Consider the program in Figure~\ref{fig:buffers}.
%
%Since the writes in the first thread are separated by a fence, the assertion
%holds under SC, TSO, PSO, and RMO.  
%
The assertion holds under SC, TSO, PSO, and RMO.
But, removing the fence causes it to fail under PSO and RMO.
In this section, we show why MM-oblivious methods may generate bogus
errors, why SC-specific ones may generate bogus proofs, and how our
new method fixes both issues.

\begin{figure}
\centering

{\footnotesize

\begin{minipage}{0.425\linewidth}
\begin{lstlisting}[language=C,frame=single]
void thread1() {
  x = 5;
  fence;
  y = 10;
}
\end{lstlisting}
\end{minipage}
\hspace{0.10\linewidth}
\begin{minipage}{0.425\linewidth}
\begin{lstlisting}[language=C,frame=single]
void thread2() {
  if (y == 10) {
    assert(x == 5);
  }
}
\end{lstlisting}
\end{minipage}
}

\caption{The assertion always holds, but if the fence is removed, the
  assertion may fail under PSO and RMO.}
\label{fig:buffers}
\vspace{-2ex}
\end{figure}

\subsection{Behaviors under SC, TSO, PSO, and RMO}

%First, we explain why the assertion holds when the program in
%Figure~\ref{fig:buffers} runs under SC.  Recall that under SC each
%thread executes its instructions in the program order, i.e.,
%\textcode{x = 5} \emph{takes effect} before \textcode{y = 10}.
%
First, note that the assertion in Figure~\ref{fig:buffers} holds under SC since
each thread executes its instructions in program order, i.e., \textcode{x = 5}
\emph{takes effect} before \textcode{y = 10}.
So, thread two observing \textcode{y} to be 10 implies \textcode{x}
must have been set to 5.

Next, we explain why the assertion holds under TSO~\cite{AdveG96}.  
%
%TSO allows the reordering of stores and loads meaning, within a thread, a store
%may be delayed after a subsequent load if the two instructions access different
%addresses (as in Figure~\ref{fig:dekker}).  
%
TSO permits the delay of a store after a subsequent load to a disjoint memory
address (as in Figure~\ref{fig:dekker}).
This \emph{program-order relaxation} is a performance optimization, e.g.,
buffering slow stores to speed up subsequent loads.
However, since all stores in a thread go into the same buffer, TSO does not
allow the reordering of two stores (thread 1, Figure~\ref{fig:buffers}).
Thus, even without the fence, \textcode{x = 5} always \emph{takes effect}
before \textcode{y = 10}, meaning the assertion holds.

%Next, we explain why the assertion fails under PSO or RMO if
%the fence is removed.  
%
Next, we show why removing the fence causes the assertion to fail under PSO and
RMO.
Both permit store--store reordering by allowing each processor to
have a separate store buffer for each memory address. 
%
%Thus \textcode{x} is in one store buffer while \textcode{y} is in another.  
%
Thus \textcode{x} and \textcode{y} are in separate buffers.
Since buffers are flushed to memory independently, with the fence
removed, \textcode{y = 10} may take effect before \textcode{x = 5}, as
if the two instructions were reordered in this thread.
Thus, the second thread may read 10 from \textcode{y} before 5 is written to
\textcode{x} in global memory, thus causing the assertion to fail.
%
%With the fence removed from the first thread, the second thread may actually
%read 10 from \textcode{y} before 5 is written to \textcode{x} in the global
%memory, thus failing the assertion.

The fence is important because it forces all stores issued before the
fence to be visible to all loads issued after the fence, i.e.,
\textcode{x = 5} takes effect before \textcode{y = 10}, even under PSO
and RMO.  Thus, the assertion holds again.

\subsection{Ineffectiveness of Existing Methods}

MM-oblivious methods~\cite{Mine11, Mine12, Mine14} report bogus alarms because
they were not designed for weak memory, and they ignore the causality
of inter-thread data flows.  
Thus, they tend to drastically over-approximate the
interferences between threads.

\begin{figure*}[t!]
\centering
\resizebox{\linewidth}{!}{%
\begin{tabular}{|l||ccc|ccc||ccc|ccc|}\hline
             & \multicolumn{6}{|c||}{Program in Figure~\ref{fig:dekker}}
             & \multicolumn{6}{|c|}{Program in Figure~\ref{fig:buffers}} \\
  \cline{2-13}  
 Method      & \multicolumn{3}{|c }{with fences}  
             & \multicolumn{3}{|c||}{without fences} 
             & \multicolumn{3}{|c }{with fences} 
             & \multicolumn{3}{|c|}{without fences} \\
  \cline{2-13}  
             &~~~SC~~~&  TSO   &  PSO/RMO   &~~~SC~~~&  TSO   &  PSO/RMO   &~~~SC~~~&  TSO   &  PSO/RMO   &~~~SC~~~&  TSO   &  PSO/RMO   \\
    \hline\hline
             &~~(bogus)~ &~~(bogus)~ &(bogus) &~~(bogus)~ &        &        &~~(bogus)~ &~~(bogus)~ &(bogus) &~~(bogus)~ &~~(bogus)~ &        \\ 
    MM-oblivious (e.g.~\cite{Mine11,Mine12,Mine14})
             &alarm   &alarm   &alarm     &alarm &{\bf alarm} &{\bf alarm}   &alarm &alarm &alarm   &alarm &alarm &{\bf alarm}   \\
    \hline
             &        &(bogus) &(bogus) &        &(bogus) &(bogus) &        &(bogus) &(bogus) &        &(bogus) &(bogus) \\
    SC-specific (e.g.~\cite{KusanoW16,FarzanK12,FarzanK13})
             &{\bf proof} &proof   &proof   &{\bf proof} &proof   &proof   &{\bf proof} &proof   &proof   &{\bf proof} &proof   &proof   \\

    \hline
             &        &        &        &        &        &        &        &        &        &        &        &        \\
    Our method
             &{\bf proof}  &{\bf proof}  &{\bf proof}  &{\bf proof}  &{\bf alarm}   &{\bf alarm}   &{\bf proof}  &{\bf proof}  &{\bf proof}  &{\bf proof}  &{\bf proof}  &{\bf alarm}   \\
    \hline
  \end{tabular}
}%resizebox
\caption{Comparing the effectiveness of various methods in handling the example programs in Figure~\ref{fig:dekker} and Figure~\ref{fig:buffers}.}
\label{fig:difference}
\end{figure*}

For example, an MM-oblivious static analysis may work as follows.
First, it analyzes each thread as if it were a sequential program.
Then, it joins the effects of all stores on global memory---known as the
\emph{thread interferences}.
Next, it individually analyzes each thread again, this time in
the presence of the thread interferences computed from the previous
iteration: when a thread performs a memory read, the value
may come from any one of these thread interferences.
This iterative process repeats until a fixed point is reached.

Next, we demonstrate how the MM-oblivious analyzer works on Figure~\ref{fig:buffers}.
Consider the thread interferences to be a map from variables to abstract values
in the interval domain~\cite{Cousot77}.  
Thread 1 generates interferences $\textcode{x} \mapsto [5,5]$ and $\textcode{y}
\mapsto [10,10]$.  
Within thread 2, the load of \textcode{y} may read
from local memory, $[0,0]$,  or the interference $[10,10]$.  
Thus, $\textcode{y} = [0,0] \sqcup [10,10] = [0,10]$, where $\sqcup$ is
the \emph{join} operator in the interval domain.
%
%So, the first then-branch can be taken.
%
Similarly, the load of \textcode{x} may read from local memory, $[0,0]$, or the
interference $[5,5]$, i.e., $\textcode{x} = [0, 5]$. Thus, the assertion is
incorrectly reported as violated.

While our previous example used the \emph{non-relational} interval domain the bogus
alarm remains when using a relational abstract domain: the propagation of
interferences in MM-oblivious methods is inherently non-relational. Inferences
map variables to a single values, causing all relations to be
forgotten.
%
% This is hard to fix using any conventional approach~\cite{Mine14}, since
% maintaining relational invariants at all global program points is prohibitively
% expensive.
%
Conventional approaches cannot easily fix this since maintaining relational
invariants at all global program points is prohibitively expensive.

In contrast, prior SC-specific
methods~\cite{KusanoW16,FarzanK12,FarzanK13} do not report bogus
alarms: they assume \textcode{x = 5} takes effect before
\textcode{y = 10}.  
This leads to more accurate analysis results for SC, but is unsound under weak
memory, e.g., they miss the assertion failure in Figure~\ref{fig:buffers} under
PSO or RMO when the fence is removed.
%
% Also, these prior methods may
% return correctness proofs, such proofs cannot be trusted under a weak
% memory model, because only SC compatible program behaviors are
% considered during the analysis.
%
%Overall, prior SC-specific methods are unsound under weak-memory since only SC
%compatible program behaviors are considered

Figure~\ref{fig:difference} summarizes the ineffectiveness of prior techniques
on the programs in Figures~\ref{fig:dekker} and \ref{fig:buffers} with and
without fences. 
Note that in Figure~\ref{fig:dekker} the fence instruction may be added
between the write and read instructions of both threads.
The table in Figure~\ref{fig:difference} shows how prior MM-oblivious methods report bogus alarms, prior
SC-specific methods report bogus proofs, while our new method eliminates both.

\subsection{How Our Method Handles Memory Models}

%The main reason why prior techniques lead to bogus alarms is because
%of their over-approximated handling of thread interferences.  
%
Some prior techniques lead to bogus alarms because they
over-approximate thread interferences, i.e., they allow a load to read
from any remote store regardless of whether such a data flow, or
combination of flows, is feasible, while others lead to missed bugs
because they under-approximate thread interferences, i.e., they do not
allow any non-SC data flow.
Consider Figure~\ref{fig:buffers}: the load
of \textcode{x} may read 0 or 5, and the load of \textcode{y} may read 0
or 10, but the combination of \textcode{x} reading 0 and \textcode{y}
reading 10 is infeasible.  
Realizing this, our method
checks the feasibility of interference combinations under weak-memory
semantics before propagating them.

Toward this end, we propose two new techniques.
The first is the flow-sensitive propagation of thread interferences.
Instead of eagerly joining all interfering stores, we handle each
combination separately.  
The second is a declarative modeling of the memory consistency semantics general enough to
capture SC, TSO, PSO, and RMO~\cite{AdveG96,Weaver94,Sites92}.
Together, these techniques prune infeasible combinations of thread interferences such as \textcode{x}
and y reading 0 and 10, respectively, in Figure~\ref{fig:buffers}.
%
%As a result, our new method is not only thread-modular, but also
%flow-sensitive, and can robustly handle a wide-range of
%processor-level memory models.

%Specifically, when analyzing the second thread in Figure~\ref{fig:buffers}, our
%method would consider four different interference combinations, denoted
%$\rho_1-\rho_4$, separately.  
%
Our new method analyzes thread 2 in Figure~\ref{fig:buffers} by
considering four different interference combinations, $\rho_1$--$\rho_4$,
separately.
%
%They correspond to the load of values from either local or global memory.
%
\begin{itemize*}
\item $\rho_1 = \textcode{y} \mapsto [0,0]$~~ and $\textcode{x} \mapsto [0,0]$,
\item $\rho_2 = \textcode{y} \mapsto [10,10]$ and $\textcode{x} \mapsto [5,5]$,
\item $\rho_3 = \textcode{y} \mapsto [0,0]$~~ and $\textcode{x} \mapsto [5,5]$,
\item $\rho_4 = \textcode{y} \mapsto [10,10]$ and $\textcode{x} \mapsto [0,0]$.
\end{itemize*}
We gain accuracy in two ways. First, we remove spurious values caused by an
eager join (e.g., we no longer have $\textcode{y} = [0, 10]$). 
Second, we query
a lightweight constraint system to quickly deduce infeasibility of an
interference combination on demand. $\rho_1$, $\rho_2$, and $\rho_3$ are all
feasible but they do not cause assertion failures.

Our check for infeasibility of an interference combination is
implemented using Datalog (Horn clauses within finite
domains), solvable in polynomial time. 
We will provides details of this constraint system in
Section~\ref{sec:feasibility}.
For now, consider $\rho_4$ in Figure~\ref{fig:buffers}: it is infeasible (unless we assume the program runs under PSO or
RMO with the fence removed).  
We deduce infeasibility as follows:
%Below is how we deduce that ``reading
%$\textcode{y} = [10,10]$ as well as $\textcode{x} = [0, 0]$'' violates
%the ordering constraints of the program:
%
%\vspace{1ex}
\begin{itemize}
\item \textcode{y = 10} has executed (it is being read from),
\item thus \textcode{x = 5} has executed (due to the \emph{program-order}
  requirement on SC and TSO, and the fence on PSO and RMO), 
\item so the load of $\textcode{x}$ must not read from its initial
  value $[0,0]$.
\end{itemize}
%\vspace{1ex}

\noindent
This deduction leads to a formal proof that $\rho_4$ can not exist in any
concrete execution.  Since the combinations $\rho_{1-3}$ do not
violate the assertion, and $\rho_4$ is proved to be infeasible, the property is verified.

\section{Preliminaries}
\label{sec:background}

In this section, we review weak memory models at the processor level
(as opposed to the programming language level) and static program
analysis based on abstract interpretation.

\subsection{Concurrent Programs}

We are concerned with a program consisting of a finite set of threads. 
Each thread assesses a set $\{a,b,c,\ldots\}$ of local variables.  
All threads access a set $\{x,y,z,\ldots\}$ of global variables via \emph{load} and \emph{store} instructions.
A thread creates a child thread with  \emph{ThreadCreate}, and waits it to
terminate with \emph{ThreadJoin}. 

We represent a program using a set $\GraphSet = \{G_1,\ldots,G_k\}$ of
flow graphs.  Each flow graph $G\in\GraphSet$, where $G = \langle N, n_0,
\delta \rangle$, is a thread: $N \subseteq \NodeSet$ is the set of
program locations of the thread, $n_0\in N$ is the entry point, and
$\delta$ is the transition relation.  That is, $(n', n) \in \delta$
iff there exists an edge from $n'$ to $n$.
%, i.e., executing the
%instruction in $n'$ may lead the control to $n$. 

\begin{figure*}
\centering
\scalebox{0.76}{
\setlength{\tabcolsep}{6pt} 
\begin{tabular}{|l|cccc|cccc|c|}\hline

             & \multicolumn{8}{|c}{Which Program-Order Relaxation Is Allowed?}  & \multicolumn{1}{|c|}{~~Write-Atomicity~~} 
\\\cline{2-10}
Memory Model~~~~~~
             & R($v_1$)$\rightarrow$R($v_1$) & R($v_1$)$\rightarrow$W($v_1$) & W($v_1$)$\rightarrow$R($v_1$)     & W($v_1$)$\rightarrow$W($v_1$)   
             & R($v_1$)$\rightarrow$R($v_2$) & R($v_1$)$\rightarrow$W($v_2$) & W($v_1$)$\rightarrow$R($v_2$)     & W($v_1$)$\rightarrow$W($v_2$)   
             & read own  
\\
             & & & & & & & & &   ~~~write early~~~ 
\\\hline\hline
%              L-L    L-S    S-L    S-S      L-L    L-S    S-L    S-S     own   
SC~\cite{Lamport79}
            &  no  &  no  &  no   &  no    &  no  &  no  &  no  &  no    & no   \\\hline    
TSO~\cite{Weaver94,Sewell10}
            &  no  &  no  &  no*  &  no    &  no  &  no  &  yes &  no    & yes   \\\hline    
PSO~\cite{Weaver94}
            &  no  &  no  &  no*  &  no    &  no  &  no  &  yes &  yes   & yes   \\\hline    
RMO~\cite{Weaver94,Sites92}~
            &  no  &  no  &  no*  &  no    &  yes &  yes &  yes &  yes   & yes   \\\hline
%PowerPC~\cite{MaySSW94}
%            &  no  &  no  &  no  &  no    &  yes &  yes &  yes &  yes   & yes   \\\hline
\end{tabular}
}
\caption{Allowed relaxations of various processor-level memory models (cf.~\cite{AdveG96}).
  $v_1$ and $v_2$ are distinct variables, and * indicates rule needs to be
  relaxed to allow \emph{read-own-write-early} behaviors (see Section~\ref{sec:write-atom} for explanation).}
\label{fig:wmm}
\end{figure*}

Each program location $n\in\NodeSet$ is associated with an atomic instruction that
may be a \emph{load}, \emph{store}, or \emph{fence}.
Non-atomic statements such as \textcode{y = x+1},
where both \textcode{x} and \textcode{y} are global variables, can be
transformed to a sequence of atomic instructions, e.g., the load
\textcode{a = x} followed by the store \textcode{y = a+1}, where
\textcode{a} is a local variable in both cases.
When accessing variables on the global memory, threads may use a
special \emph{fence} instruction to impose a strict program order
between memory operations issued before and after the fence.
%
%Since in this work, we target programs written using POSIX threads, we
%assume programmers may implement thread-level concurrency control
%using primitives such as mutex locks (\textcode{lock/unlock}) and
%condition variables (\textcode{signal/wait}).  
%%
%These synchronization primitives actually use \emph{fence} instructions
%internally, so they will also impose the strict program order between memory
%operations.

\subsection{Memory Consistency Models}

The simplest memory model is sequential consistency
(SC)~\cite{Lamport79}. 
SC corresponds to a system running on a single coherent memory time-shared by
operations executed from different threads.  
%
%This is equivalent to the interleaved execution of operations from multiple
%threads by a uniprocessor.
%
There are two important characteristics of SC: the
\emph{program-order} requirement and the \emph{write-atomicity}
requirement.
The program-order requirement says that the processor must
ensure that instructions within a thread take effect in the order
they appear in the program.
The write-atomicity requirement says that the processor must maintain the
illusion of a single sequential order among operations from all threads. That
is, the effect of any store operation must take effect and become visible
either to \emph{all} threads or to \emph{none} of the threads.

SC is an ideal memory model: In real CPUs, the hardware-level memory models are often weaker than SC, and
can be characterized by their corresponding relaxations of the \emph{program-order}
and \emph{write-atomicity} requirements as shown in Figure~\ref{fig:wmm}.  
Here, $R(v_1)\rightarrow W(v_2)$ is a read of $v_1$ followed by a write of
$v_2$ in the same thread.

Specifically, TSO allows \textcode{x=1;a=y} to be reordered as
\textcode{a=y;x=1}, according to W($v_1$)$\rightarrow$R($v_2$) in
Column 8, where $v_1$ is \textcode{x} and $v_2$ is \textcode{y}.
PSO further allows \textcode{x=1;y=2} reordered to \textcode{y=2;x=1},
according to W($v_1$)$\rightarrow$W($v_2$) in Column 9.
As shown in Section~\ref{sec:motivation}, these program-order relaxations,
conceptually, are the effect of store buffering, which delay the stores past subsequent
stores/loads within a thread.
Neither TSO nor PSO permits the delay of a load.
Weaker still is RMO, which  permits the relaxations of R($v_1$)$\rightarrow$R($v_2$)
and R($v_1$)$\rightarrow$W($v_2$), as shown in Columns 6 and 7 of the table in Figure~\ref{fig:wmm}. 

%However, while discussing the program-order relaxations, it is
%important to note that multi-core processors need to maintain backward
%compatibility for sequential software written for uniprocessors.  For
%example, in the code snippet \textcode{x=1; a=x; if(a==1){...}}, since
%%the store \textcode{x=1} and the subsequent load \textcode{a=x} access
%the same variable \textcode{x}, they cannot be reordered---otherwise,
%the semantics of the program will change---as indicated by
%Columns~2--5 of the table in Figure~\ref{fig:wmm}.

By relaxing the write-atomicity requirement, all three weaker
memory models allow a thread to read its own write early.  That is,
the thread can read a value it has written before the value reaches
the global memory and hence becomes visible to other threads.

\subsection{Abstract Interpretation}

Abstract interpretation is a popular technique for conducting static
program analysis~\cite{Cousot77}.  In this context, a numerical
abstract domain defines, for every $n\in\NodeSet$ of the program, a
memory environment $s$.  It is a map from each program variable to its
abstract value\footnote{For ease of presentation we assume a variable maps to a
single value. Our analysis can trivially use relational domains.}.
Consider intervals, which map each variable to a region defined
by the lower and upper bounds.  For a program with two integer
variables $x$ and $y$ where both may have any value initially, the
memory environment associated with the entry point $n_0\in\NodeSet$ is
$s_0 = \{x \mapsto \top, y \mapsto \top\}$, where $\top =
(-\infty,+\infty)$.  After executing \textcode{x = 1}, the memory
environment becomes $s_1 = \{x \mapsto [1,1], y \mapsto \top\}$.

The process of computing $s_1$ based on $s_0$ is represented by the
transfer function of \textcode{x = 1}.  Additionally, the join is
defined as $[l_1,u_1] \sqcup [l_2,u_2] = [\mathit{min}(l_1,l_2),
  \mathit{max}(u_1,u_2)]$.  The partial-order relation is defined as
$[l_1,u_1] \sqsubseteq [l_2,u_2]$ if and only if $l_1\geq l_2$ and
$u_1\leq u_2$.  For example $[1,3] \sqcup [7,10] = [1,10]$ and $[4, 6]
\sqsubseteq [1, 10]$.

We use $\StateSet$ to denote the set of all memory environments.
$\StateSet$ is a lattice with properly defined top ($\top$) and bottom
($\bot$) elements, join ($\sqcup$), partial-order ($\sqsubseteq$), and
a widening operator~\cite{Cousot77}.
Each node $n\in\NodeSet$ has a \emph{transfer function} $t \in
\StateSet \rightarrow \StateSet$, taking an environment
$s'\in\StateSet$ as input (before executing the atomic operation in
$n$) and returns a new environment $s\in\StateSet$ as output.

Let $\tfunc \in \NodeSet \rightarrow (\StateSet \rightarrow
\StateSet)$ be a map from each node to its transfer function.
For example, given a node $n\in\NodeSet$ whose operation is
\textcode{x = a+1}, if $s' = \{x \mapsto [1,3], a \mapsto [2,5]\}$,
the new environment is $s = (\tfunc(n))(s') = \{x \mapsto [3,6], a
\mapsto [2,5]\}$.

The goal of an abstract interpreter is to compute an environment map
$M \in (\NodeSet \rightarrow \StateSet)$ over-approximating the
memory state at every program location.
$M$, typically, initially maps all variables in the entry
node to $\top$, and all variables in other nodes to
$\bot$.  Then, it iteratively applies the transfer function
$\tfunc(n)$ and joins the resulting environments for all $n$, until
they reach a fixed point.

Without getting into more details (refer to the
literature~\cite{Cousot77}), we define the sequential analyzer as a
fixed-point computation with respect to the function $\AnalyzeSeq \in
(\NodeSet \rightarrow \StateSet) \rightarrow (\NodeSet \rightarrow
\StateSet)$:
\begin{align*}
  \AnalyzeSeq(M) = n \mapsto (\tfunc(n))(\bigsqcup_{(n',n) \in \delta}
  M(n'))
\end{align*}
Here, $M(n')$ is the environment produced by a predecessor node $n'$
of $n$, and $s'=\bigsqcup_{(n',n) \in \delta} M(n')$ is the join of
these environments.  $(\tfunc(n))(s')$ is the new memory environment
produced by executing the operation in $n$.
Applying this function to all nodes of a sequential program until a
fixed point leads to an over-approximated memory state for each
program location.

%\subsection{Thread-Modular Abstract Interpretation}

However, directly applying the sequential analyzer to each execution
of a multithreaded program is not practical because it leads to an
exponential complexity.
Instead, thread-modular techniques~\cite{Mine11, Mine12, Mine14, KusanoW16}
iteratively apply $\AnalyzeSeq$ to each thread, as if it were a sequential
program, and then merge/propagate the global memory effects across threads.
The iterative process continues until memory environments in all threads
stabilize.

%Since abstract interpretation is carried out on each thread in
%isolation in each iteration, this approach is more scalable than
%non-thread-modular techniques.  
%
Since each thread is analyzed in isolation, this approach is more scalable than
non-thread modular techniques.
However, it may result in
accuracy loss because the analyzer for each thread relies on a
coarse-grained abstraction of \emph{interferences} from other threads.
When analyzing a thread $t$ in the presence of a set of threads
$T$, for example, the interferences are the effects of global memory stores from
all $t' \in T$.
The interferences are a map $(\VarSet \rightarrow
2^{\StateSet})$ from each variable $v\in\VarSet$ read by thread $t$ to
the set of memory environments produced by interfering stores, where
$\VarSet$ is the set of all program variables, and $2^{\StateSet}$ is the power set of $\StateSet$.

Prior thread-modular techniques~\cite{Mine11, Mine12, Mine14} eagerly
join all interfering memory states from the other threads in $T$
before propagating them to the current thread $t$.  As such, they
often introduce bogus \emph{store-to-load} data flows into the static
analysis or miss valid \emph{store-to-load} data flows.  In the
remainder of this paper, we present our method for mitigating this
problem.

%\clearpage
%\newpage
%\section{Deciding the Feasibility of Interferences on Relaxed Memory}
\section{Deciding Interference Feasibility}
\label{sec:feasibility}

In this section, we describe our new method for quickly deciding the
feasibility of a combination of store-to-load data-flows under a given memory
model.
An \emph{interference combination} is a set $\mathit{ic} = \{ (l, s),
\ldots \}$ where each $(l,s) \in \mathit{ic}$ is a load $l$ and an
interfering store $s$.

Checking the feasibility of $ic$ is formulated as a deductive analysis
with inputs: (1) the flow graph of the current thread, (2) the flow
graphs of all interfering threads, and (3) the existing set of
store-to-load data flows represented by $\{(l,s)|
(l,s)\in\textsc{ReadsFrom}\}$.  The output of this deductive analysis
is the relation \textsc{MustNotReadFrom}.
$(l,s)\in\textsc{MustNotReadFrom}$ means the load $l$ must not read from the
store $s$ since our analysis proved the data flow from $s$ to $l$ is infeasible
given the input \textsc{ReadsFrom} relation in $ic$.

Consider the program in Figure~\ref{fig:buffers} as an example. One thread
interference combination we want to check is the load of \textcode{y}
from \textcode{y=10} and the load of \textcode{x} from the initial
value 0.  Let these load and store instructions be denoted $l_y$,
$s^{10}_y$, $l_x$, and $s^{0}_x$, respectively.  Then, the feasibility
problem is stated as follows: given $(l_y,s^{10}_y) \in
\textsc{ReadsFrom}$, check if
$(l_x,s^{0}_x)\in\textsc{MustNotReadFrom}$.

\subsection{Notations}

Before presenting the details of our feasibility checking procedure,
we define a set of unary and binary relations over instructions and
program variables.  
Specifically, $(s_1,v_1)\in\textsc{IsLoad}$
denotes $s_1$ is a load of variable $v_1$, and
$(s_2,v_2)\in\textsc{IsStore}$ denotes $s_2$ is a store to
variable $v_2$.  We use $(s_1,\_)\in\textsc{IsLoad}$
if we do not care about the variable.
Similarly, we use $f\in\textsc{IsFence}$ to denote that
$f$ is a fence.  
We also use $\textsc{IsLLMembar}$, $\textsc{IsLSMembar}$,
$\textsc{IsSLMembar}$, $\textsc{IsSSMembar}$ to denote load--load,
load--store, store--load, and store--store memory barriers as defined
in the SPARC architecture~\cite{Weaver94}; for example, a load--store
\texttt{membar} prevents loads before the barrier from being reordered
with subsequent stores.

We define binary relations over instructions $s_1$ and $s_2$:
%
%\noindent
%
the first four relations (\textsc{Dominates, NotReachableFrom,
  ThreadCreates, ThreadJoins}) are determined by the program's flow
graphs. Based on them, we deduce the \textsc{MHB} relation, which must
be satisfied by all program executions.  The \textsc{ReadsFrom}
relation comes from the given $ic$, from which we deduce the
\textsc{MustNotReadFrom} relation.

\vspace{2ex}
\noindent
{\footnotesize
\begin{tabular}{lp{.745\linewidth}}
\toprule

$(s_1,s_2)\in$ & $\textsc{Dominates}$~~ means that $s_1$ dominates $s_2$
in the control flow graph of a thread.\\

$(s_1,s_2)\in$ & $\textsc{NotReachableFrom}$~~ means that $s_1$ cannot
be reached from $s_2$ in the control flow graph of a thread.\\

$(s_1,s_2)\in$ & $\textsc{ThreadCreates}$~~ means $s_1$ is the thread
creation and $s_2$ is the first operation of the child thread.\\

$(s_1,s_2)\in$ & $\textsc{ThreadJoins}$~~ means $s_1$ is the thread join
and $s_2$ is the last operation of the child thread.\\

$(s_1,s_2)\in$ & $\textsc{MHB}$~~ means that $s_1$ must happen before
$s_2$ in all executions of the program.\\

$(s_1,s_2)\in$ & $\textsc{ReadsFrom}$~~ means that $s_1$ is a load
that reads the value written by the store $s_2$.\\

$(s_1,s_2)\in$ & $\textsc{MustNotReadFrom}$~~ means that $s_1$ must not
read from the value written by $s_2$.\\ \bottomrule
\end{tabular}
}
\vspace{2ex}

Consider Figure~\ref{fig:buffers} again, where we want to check if
the load of \textcode{y} in the second thread reads from
\textcode{y=10}, then is it possible for the load of \textcode{x} to
read from the initial value 0?
In this case, we encode the assumption as $(l_y,s^{10}_y) \in
\textsc{ReadsFrom}$.  Next, we deduce the \textsc{MustNotReadFrom}
relation.  Finally, we check if
$(l_x,s^{0}_x)\in\textsc{MustNotReadFrom}$.

\subsection{Relaxing the Program-Order Requirement}
\label{sec:program-order}

To model the program order imposed by different memory models, we
define a new relation $\textsc{NoReorder}$ such that
$(s_1,s_2)\in\textsc{NoReorder}$ if the reordering of $s_1$ and $s_2$
within the same thread is not allowed.

%Based on how the program-order requirement is relaxed in different
%memory models, as shown in Figure~\ref{fig:wmm}, we define the rules
%for deducing \textsc{NoReorder}.
%
We define the rules for \textsc{NoReorder} based on the allowed program-order
relaxations for different memory models (Figure~\ref{fig:wmm}).

For SC, \textsc{NoReorder} is defined as:
{\footnotesize
\begin{equation*}
  \label{eqn:NoReorder-SC}
  \begin{mathprooftree}
    \AxiomC{ $\top$ }
    \RightLabel{ (under SC) }
    \UnaryInfC{ $(s_1,s_2)\in\textsc{NoReorder}$ }
  \end{mathprooftree}
\end{equation*}  
}
%
%That is, no reordering is ever allowed under SC, as shown by Columns 2-9 of the
%table in Figure~\ref{fig:wmm}.
That is, no reordering is ever allowed under SC (row SC Figure~\ref{fig:wmm}).
%table in Figure~\ref{fig:wmm}.

For TSO, \textsc{NoReorder} is defined as:
{\footnotesize
\begin{equation*}
  \label{eqn:NoReorder-TSO}
  \begin{mathprooftree}
    \AxiomC{ $(s_1,\_)\in\textsc{IsLoad}$ } 
    \RightLabel{ (under TSO) }
    \UnaryInfC{ $(s_1,s_2)\in\textsc{NoReorder}$ }
  \end{mathprooftree}\\
\end{equation*}  
\begin{equation*}
  \begin{mathprooftree}
    \AxiomC{ $(s_2,\_)\in\textsc{IsStore}$ } 
    \RightLabel{ (under TSO) }
    \UnaryInfC{ $(s_1,s_2)\in\textsc{NoReorder}$ }
  \end{mathprooftree}\\
\end{equation*}%
}%
Under TSO, two operations $(s_1,s_2)$ can not reorder in six of the eight
cases.
The first rule above disallows Columns 2, 3, 6, and 7 (Figure~\ref{fig:wmm}),
while the second disallows Columns 3, 5, 7, and 9.  
%
%Thus, the only two not-yet-covered cases (where reordering is allowed) are
%Columns 4 and 8. 
%
Thus, reordering is permitted in two cases: Columns 4 and 8.

%Note that while TSO does not allow a write and a read of the same variable to
%be reordered (Column 4), it must be allowed to ensure soundness; the same holds
%for PSO and RMO. We discuss this shortly in Section~\ref{sec:write-atom}.
%
Although this is counter-intuitive, note that $W(v_1)\rightarrow
R(v_1)$ (Column 4) may be reordered in our analysis under TSO (and PSO
and RMO) for soundness: it permits \emph{read-own-write-early}
behaviors.
%
%The detailed explanation is in Section~\ref{sec:write-atom}
%
We detail this shortly in Section~\ref{sec:write-atom}.

For PSO, \textsc{NoReorder} is defined as:
{\footnotesize
\begin{equation*}
    \label{eqn:NoReorder-PSO}
  \begin{mathprooftree}
    \AxiomC{ $(s_1,\_)\in\textsc{IsLoad} $ } 
    \RightLabel{ (under PSO) }
    \UnaryInfC{ $(s_1,s_2)\in\textsc{NoReorder}$ }
  \end{mathprooftree}
\end{equation*}  
\begin{equation*}
  \begin{mathprooftree}
   \AxiomC{ $(s_1,v_1)\in\textsc{IsStore} \land (s_2,v_1)\in\textsc{IsStore}$ } 
    \RightLabel{ (under PSO) }
    \UnaryInfC{ $(s_1,s_2)\in\textsc{NoReorder}$ }
  \end{mathprooftree}
\end{equation*}%
}%
Under PSO, two operations $(s_1,s_2)$ can not reorder in five of the eight
cases. 
The first rule above disallows Columns 2, 3, 6, and 7, while the
second disallows Column 5.
%
%The remaining three cases, permitting reordering, are Columns 4, 8, and 9.
%
Thus, reordering is permitted only in the remaining three cases (Columns 4, 8, and
9).

For RMO, the inference rules are defined as:
{\footnotesize%
  \begin{equation*}
    \label{eqn:NoReorder-RMO}
  \begin{mathprooftree}
    \AxiomC{ $(s_1,v_1)\in\textsc{IsLoad} \land (s_2,v_1)\in\textsc{IsLoad}$ } 
    \RightLabel{ (under RMO) }
    \UnaryInfC{ $(s_1,s_2)\in\textsc{NoReorder}$ }
  \end{mathprooftree}
  \end{equation*}  
  \begin{equation*}
  \begin{mathprooftree}
   \AxiomC{ $(s_1,v_1)\in\textsc{IsLoad} \land (s_2,v_1)\in\textsc{IsStore}$ } 
    \RightLabel{ (under RMO) }
    \UnaryInfC{ $(s_1,s_2)\in\textsc{NoReorder}$ }
  \end{mathprooftree}
  \end{equation*}
  \begin{equation*}
  \begin{mathprooftree}
   \AxiomC{ $(s_1,v_1)\in\textsc{IsStore} \land (s_2,v_1)\in\textsc{IsStore}$ } 
    \RightLabel{ (under RMO) }
    \UnaryInfC{ $(s_1,s_2)\in\textsc{NoReorder}$ }
  \end{mathprooftree}
  \end{equation*}%
}%
Similarly, the above inference rules can be directly translated from
Columns 2, 3, and 6 of the table in Figure~\ref{fig:wmm}.

% markus: i've moved the explanation of read-own-write early up to when it is
% first encountered with TSO. One reviewer thought we had a typo since column 4
% (the one referring to this rule) is a "no" but we say it is permitted
%The reason why we allow the reordering of $W(v_1)\rightarrow R(v_1)$,
%where write $W(v_1)$ and read $R(v_1)$ are from the same thread, is
%because it is necessary for our deduction rules to permit
%\emph{read-own-write-early} behaviors.  The detailed explanation is in
%Section~\ref{sec:write-atom}.

\subsection{Handling Fences and Memory Barriers}

%Next, we present the ordering constraints imposed by fences and memory barriers for TSO, PSO, and RMO/Alpha. 
%
Next, we present the ordering constraints imposed by fences and memory
barriers.
We consider four variants of the \texttt{membar} instruction, which
prevents loads and/or stores before the \texttt{membar} from being reordered
with subsequent loads and/or stores~\cite{Weaver94}.

{\footnotesize
\begin{equation*}
  % membar LL
  \begin{mathprooftree}
    \AxiomC{ 
      \Shortstack{
        {$m\in\textsc{IsLLMembar} \land (s_1, \_) \in \textsc{IsLoad} \land (s_2, \_) \in \textsc{IsLoad}$
        } 
        {
          $\land (s_1, m) \in \textsc{Dominates} \land (m, s_2) \in \textsc{Dominates}$
        }
      }% Shortstack
    } %AxiomC
    \UnaryInfC{ $(s_1,s_2)\in\textsc{NoReorder}$ }
  \end{mathprooftree}
\end{equation*}
\begin{equation*}
  % membar LS
  \begin{mathprooftree}
    \AxiomC{ 
      \Shortstack{
        {$m\in\textsc{IsLSMembar} \land (s_1, \_) \in \textsc{IsLoad} \land (s_2, \_) \in \textsc{IsStore}$
        }
        {
          $\land (s_1, m) \in \textsc{Dominates} \land (m, s_2) \in \textsc{Dominates}$
        }
      }% Shortstack
    } %AxiomC
    \UnaryInfC{ $(s_1,s_2)\in\textsc{NoReorder}$ }
  \end{mathprooftree}
\end{equation*}
\begin{equation*}
  % membar SL
  \begin{mathprooftree}
    \AxiomC{ 
      \Shortstack{
        {$m\in\textsc{IsSLMembar} \land (s_1, \_) \in \textsc{IsStore}  \land (s_2, \_) \in \textsc{IsLoad}$
        } 
        {
          $\land (s_1, m) \in \textsc{Dominates} \land (m, s_2) \in \textsc{Dominates}$
        }
      }% Shortstack
    } %AxiomC
    \UnaryInfC{ $(s_1,s_2)\in\textsc{NoReorder}$ }
  \end{mathprooftree}
\end{equation*}
\begin{equation*}
  % membar SS
  \begin{mathprooftree}
    \AxiomC{ 
      \Shortstack{
        {$m\in\textsc{IsSSMembar} \land (s_1, \_) \in \textsc{IsStore} \land (s_2, \_) \in \textsc{IsStore}$
        } 
        {
          $\land (s_1, m) \in \textsc{Dominates} \land (m, s_2) \in \textsc{Dominates}$
        }
      }% Shortstack
    } %AxiomC
    \UnaryInfC{ $(s_1,s_2)\in\textsc{NoReorder}$ }
  \end{mathprooftree}
\end{equation*}
}

%A fence instruction prevents the reordering of both loads and stores
%before the fence with subsequent loads and stores. This can be modeled
%in terms of \textcode{membar}s:
%
We also model fences in terms of \textcode{membar}s since they prevent loads and
stores from being reordered with subsequent loads and stores as well.

{\footnotesize
\begin{equation*}
  % membar SS
  \begin{mathprooftree}
    \AxiomC{ $f \in \textsc{IsFence}$ } 
    \UnaryInfC{ $f \in \textsc{IsLLMembar}$ }
  \end{mathprooftree}
  \hspace{8ex}
  \begin{mathprooftree}
    \AxiomC{ $f \in \textsc{IsFence}$ } 
    \UnaryInfC{ $f \in \textsc{IsLSMembar}$ }
  \end{mathprooftree}
\end{equation*}%
\begin{equation*}
  % membar SS
  \begin{mathprooftree}
    \AxiomC{ $f \in \textsc{IsFence}$ } 
    \UnaryInfC{ $f \in \textsc{IsSLMembar}$ }
  \end{mathprooftree}
  \hspace{8ex}
  \begin{mathprooftree}
    \AxiomC{ $f \in \textsc{IsFence}$ } 
    \UnaryInfC{ $f \in \textsc{IsSSMembar}$ }
  \end{mathprooftree}
\end{equation*}%
}%

In addition to fences explicitly added to the program, there are fences implicitly added to thread routines such
as \textcode{lock/unlock} and \textcode{signal/wait}.
For example, in the code snippet \textcode{x = 1; lock(lk); a = y;
  unlock(lk)}, there is a fence inside \textcode{lock(lk)}, 
to ensure \textcode{x = 1} always takes effect before \textcode{a =
  y}.
This is how most modern programming systems guarantee
data-race-freedom~\cite{AdveB10} to application-level code (i.e., programs
without data races have only SC behaviors).
Thus, we model every call $s_c$ to a POSIX thread routine using
$s_c\in\textsc{IsFence}$.

%Finally, processor-level atomic instructions such as compare-and-swap and,
%e.g., those with the \texttt{LOCK} prefix on x86~\cite{IntelGuide}~(Sec.\
%8.1.2.2), can be modeled using combinations of load, store, and fence.

\subsection{Rules for Deducing \textsc{MustNotReadFrom}}

We divide our inference rules into two groups. 
The first (Figure~\ref{fig:rules}) use the relations
\textsc{ThreadCreates}, \textsc{ThreadJoins}, \textsc{Dominates}, and
\textsc{NoReorder} to generate the must-happen-before
(\textsc{MHB}) relation.

\begin{figure}%
\begin{minipage}{\linewidth}%
{\footnotesize%
  \begin{equation}
    \label{pt:create}
    \begin{mathprooftree}
      \AxiomC{$(s, s_\mathit{sta}) \in \textsc{ThreadCreates}$}
      \UnaryInfC{$(s, s_\mathit{sta}) \in \textsc{MHB}$}
    \end{mathprooftree}
    \hspace{2ex}
    \begin{mathprooftree}
      \AxiomC{$(s, s_\mathit{end}) \in \textsc{ThreadJoins}$}
      \UnaryInfC{$(s_\mathit{end}, s) \in \textsc{MHB}$}
    \end{mathprooftree}%
  \end{equation}%
  \begin{equation}%
    \label{pt:dom}
    \begin{mathprooftree}%
      \AxiomC{
        \Shortstack{ 
          { $(s_1,s_2)\in\textsc{Dominates} \land (s_2,s_1)\in\textsc{NotReachableFrom}$ }
          { $\land$ $(s_1,s_2)\in\textsc{NoReorder}$ }
        }
      }
      \UnaryInfC{$(s_1,s_2) \in \textsc{MHB}$}
    \end{mathprooftree}%
  \end{equation}%
%
%
%  \begin{equation}
%     \label{pt:dom-i-fence}
%    \begin{mathprooftree}
%      \AxiomC{
%        \Shortstack{
%          {
%            $(s, f) \in \textsc{Dominates}
%            \land (s,f) \in \textsc{NotReachableFrom}$
%          }
%          {
%             $\phantom{e} \land f \in \textsc{IsFence}$
%          }
%        }%Shortstack
%      }%AxiomC
%      \UnaryInfC{$(s, f) \in \textsc{MHB}$}
%    \end{mathprooftree}
%  \end{equation}
%  
%  \begin{equation}
%    \label{pt:dom-fence-i}
%    \begin{mathprooftree}
%      \AxiomC{
%        \Shortstack{
%          {
%            $(f, s) \in \textsc{Dominates}
%            \land (f,s) \in \textsc{NotReachableFrom}$
%          }
%          {
%             $\phantom{e} \land f \in \textsc{IsFence}$
%          }
%        }%Shortstack
%      }%AxiomC
%      \UnaryInfC{$(f, s) \in \textsc{MHB}$}
%    \end{mathprooftree}
%  \end{equation}
% 
  \begin{equation}
    \label{pt:trans}
    \begin{mathprooftree}
      \AxiomC{
        $(s_1, s_2) \in \textsc{MHB}
         \land (s_2, s_3) \in \textsc{MHB}$
      }
      \UnaryInfC{$(s_1, s_3) \in \textsc{MHB}$}
    \end{mathprooftree}
  \end{equation}

  \begin{equation}
    \label{pt:overwrite}
    \begin{mathprooftree}
      \AxiomC{
        \Shortstack{
            {
              $(l, s_1) \in \textsc{ReadsFrom} 
              \land (s_1, s_2) \in \textsc{MHB}$
            }
            {
              $\phantom{e} \land (l, v) \in \textsc{IsLoad}
              \land (s_1, v) \in \textsc{IsStore}
              \land (s_2, v) \in \textsc{IsStore}$
            }
        } % Shortstack
      } % AxiomC
      \UnaryInfC{$(l, s_2) \in \textsc{MHB}$}
    \end{mathprooftree}
  \end{equation}%

}%
  \caption{Deduction rules for \textsc{MHB} (must-happen-before).}
  \label{fig:rules}
\end{minipage}

\vspace{4ex}

\begin{minipage}{\linewidth}
{\footnotesize

  \begin{equation}
    \label{pt:mhb_nrf}
    \begin{mathprooftree}
      \AxiomC{ $(l, s) \in \textsc{MHB}$  }
      \UnaryInfC{$(l, s) \in \textsc{MustNotReadFrom}$}
    \end{mathprooftree}
  \end{equation}

  \begin{equation}
    \label{pt:l_after_s}
    \begin{mathprooftree}
      \AxiomC{
        \Shortstack{
          {
            $(l_1, s_1) \in \textsc{ReadsFrom}
            \land (l_1, s_2) \in \textsc{MHB}
            \land (s_2, l_2) \in \textsc{MHB}$ 
          }
          {
            $\land (l_1, v) \in \textsc{IsLoad}
            \land (l_2, v) \in \textsc{IsLoad}
            \land (s_2, v) \in \textsc{IsStore}$
          }
        }
      }
      \UnaryInfC{$(l_2, s_1) \in \textsc{MustNotReadFrom}$}
    \end{mathprooftree}
  \end{equation}
}
\caption{Deduction rules for the \textsc{MustNotReadFrom}.}
\label{fig:output-rules}
\end{minipage}

\end{figure}

Rule~(\ref{pt:create}) states that if the instruction $s$ creates a 
thread with entry instruction $s_{\mathit{sta}}$, then $s$ must happen
before $s_{\mathit{sta}}$. 
Similarly, if instruction $s$ joins a
thread with exit instruction $s_{\mathit{end}}$, then
$s_{\mathit{end}}$ must happen before $s$.  
%
%The correctness of this rule is obvious.

\begin{figure}
\centering
\vspace{-2ex}
\includegraphics[width=.8\linewidth]{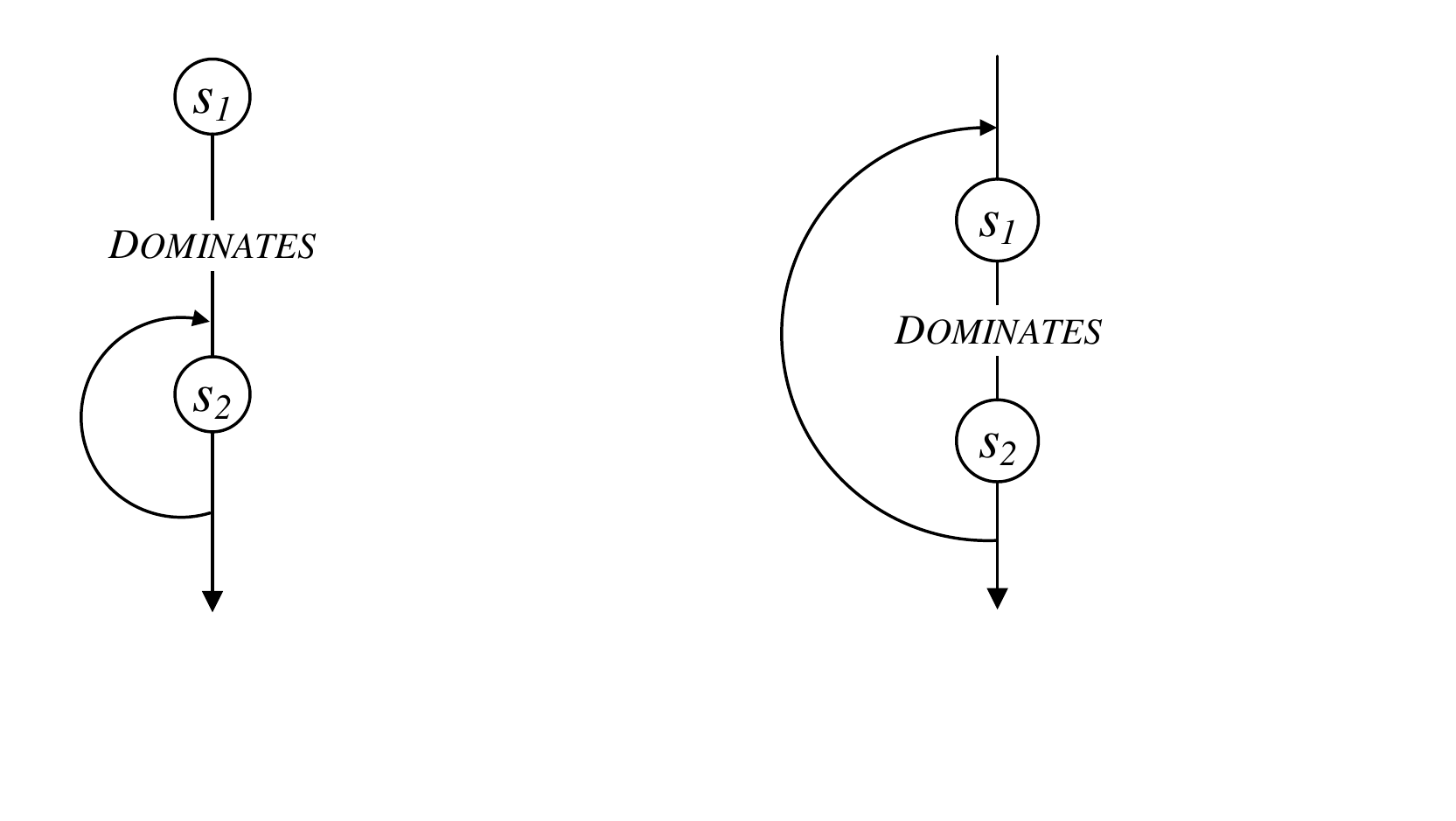}
\vspace{-6ex}
  \caption{Example illustrating Rule~(\ref{pt:dom})}
\label{fig:dom}
\vspace{-2ex}
\end{figure}

Rule~(\ref{pt:dom}) states that if $s_1$ dominates $s_2$ within a thread's
CFG, and $s_1$ is not reachable from $s_2$, (i.e., no loop encompasses both
$s_1$ and $s_2$), then, if permitted by the memory model, $s_1$ must happen
before $s_2$.
Figure~\ref{fig:dom} exemplifies this rule: the loop in
the left CFG is outside the \textsc{Dominates} edge, thus
$(s_1,s_2)\in\textsc{NotReachableFrom}$. The loop in the right CFG encompasses
the \textsc{Dominates} edge, thus $(s_1,s_2)\not\in\textsc{NotReachableFrom}$.
%
%The correctness of this rule is obvious. 

Rule~(\ref{pt:trans}) states that the \textsc{MHB} relation is transitive: if
$s_1$ must happen before $s_2$, and $s_2$ must happen before $s_3$, then $s_1$
must happen before $s_3$.  
Correctness follows from the definition of \textsc{MHB}.
%The correctness of this rule directly follows the definition of \textsc{MHB}.

\begin{figure}
\centering
\vspace{-2ex}
\includegraphics[width=0.75\linewidth]{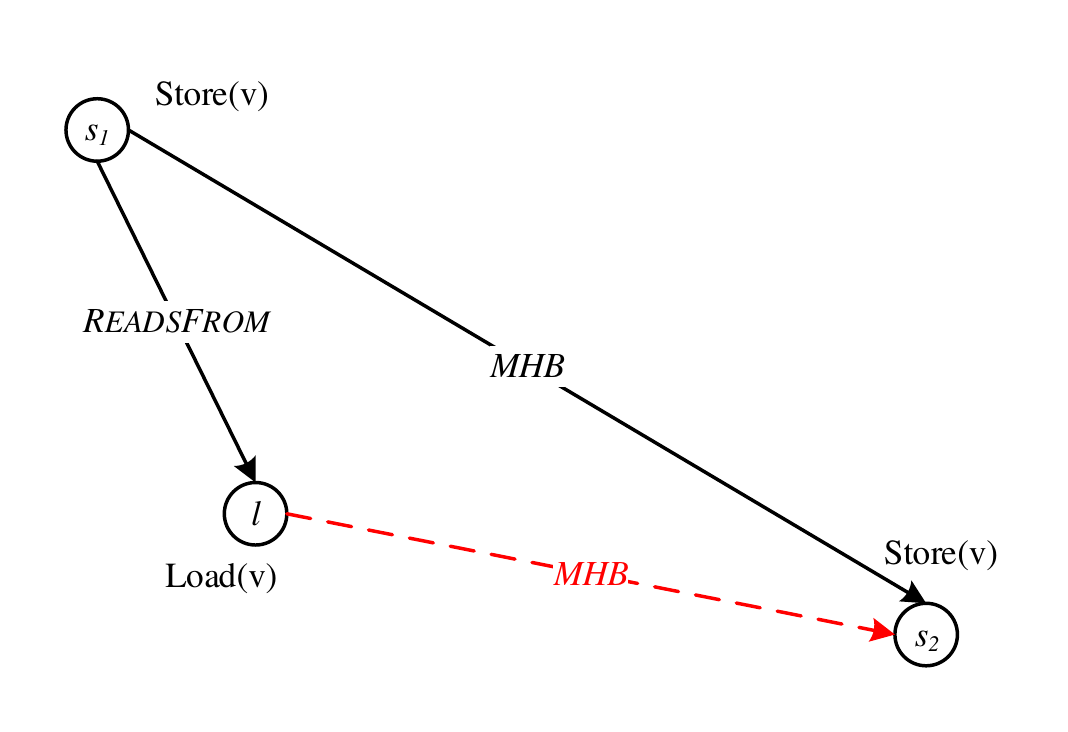}
\vspace{-2ex}
%\caption{Example for illustrating Deduction Rule~(\ref{pt:overwrite}), where $(l,s_2)\in\textsc{MHB}$ is deduced.}
  \caption{Example illustrating Rule~(\ref{pt:overwrite})}
\label{fig:overwrite}
\end{figure}

Rule~(\ref{pt:overwrite}) states that if a load $l$ reads from the
value written by the store $s_1$, then $l$ must happen before some
second store to the same variable $s_2$ takes effect.  
This is intuitive because, if $s_2$ takes effect before $l$ (but after the
first store $s_1$), then $l$ can no longer read from $s_1$.
Figure~\ref{fig:overwrite} exemplifies this rule. Its correctness is obvious.

%\vspace{1ex} 
%
The second group of inference rules (Figure~\ref{fig:output-rules}) takes the
relations \textsc{MHB} and \textsc{ReadsFrom} and generates the
\textsc{MustNotReadFrom} relation.  
Recall that if a load-store pair $(l,s) \in \textsc{MustNotReadFrom}$, the value stored by $s$ can never
flow to $l$.
Thus, \textsc{MustNotReadFrom} may be used to eliminate infeasible data flows.
%from the iterative static analysis of individual threads.

%\vspace{1ex}
Rule~(\ref{pt:mhb_nrf}) states that if a load $l$ must
happen before a store $s$, then $l$ cannot 
read from $s$.  This follows from the definition of
\textsc{MHB}.  
%
%Note that for a store $s$ to ``happen'' means when the
%store takes effect.
%
Note that a store $s$ ``happens'' when it propagates to main memory.

\begin{figure}
\centering
\vspace{-2ex}
\includegraphics[width=0.8\linewidth]{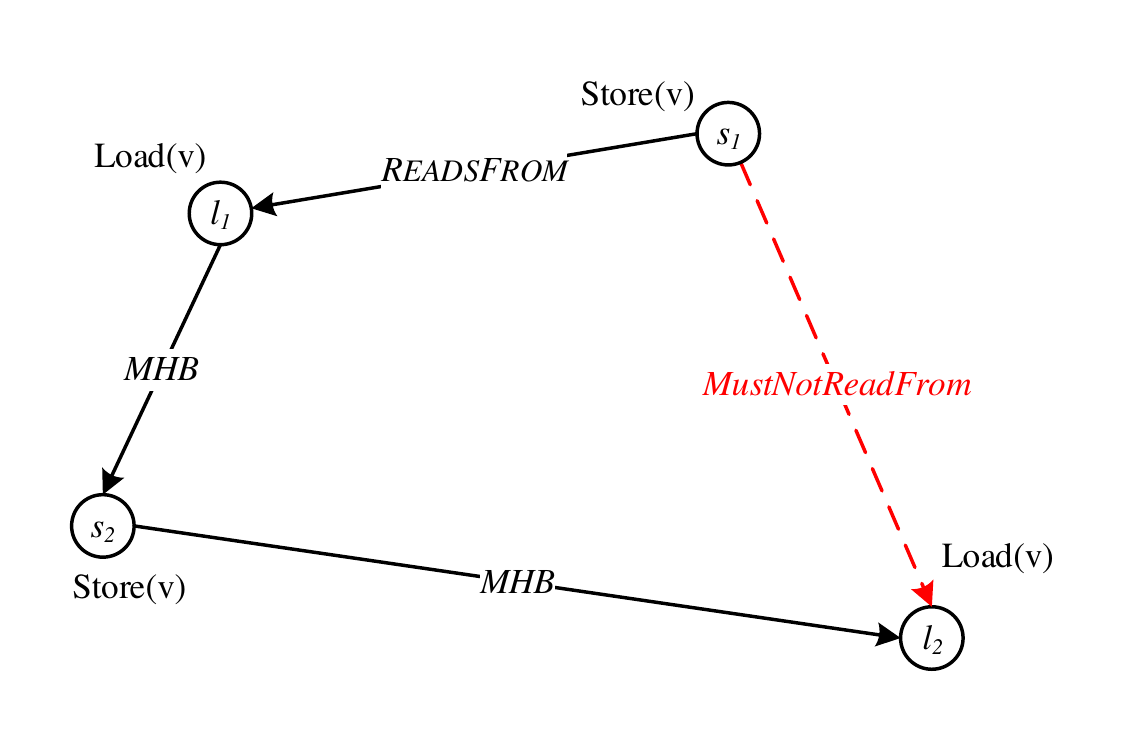}
\vspace{-2ex}
%\caption{Example for illustrating Deduction Rule~(\ref{pt:l_after_s}), where $(l_2,s_1)\in\textsc{MustNotReadFrom}$ is deduced.}
\caption{Example illustrating Rule~(\ref{pt:l_after_s}).}
\label{fig:l_after_s}
\vspace{-2ex}
\end{figure}

Rule~(\ref{pt:l_after_s}) states that if a load $l_1$ reads from a
store $s_1$, and $l_1$ must happen before some other store $s_2$, and
$s_2$ must happen before a second load $l_2$, then 
$l_2$ cannot read from $s_1$.
Figure~\ref{fig:l_after_s} exemplifies this rule.
This is correct because $l_2$ reading from $s_1$ would mean $s_1$
takes effect after $s_2$ thus preventing $l_1$ from reading $s_1$.

%\subsection{Soundness and Incompleteness of the Deduction}
\subsection{Soundness and Incompleteness}

%Our deductive analysis is sound but incomplete in terms of deciding the
%feasibility of interference combinations.  
%
When deciding the feasibility of an interference combination our analysis is
designed to be sound but incomplete.
%
%By sound, we mean it is over-approximated: it permits at least all possible
%program behaviors allowed by a memory model.
%
By sound we mean it permits all possible program behaviors allowed by
a memory model.
Therefore, if it says a certain interference combination is infeasible it must
be infeasible.
%
%However, if it does not say the interference combination is infeasible, there
%is no guarantee that it is feasible.
%
However, there is no guarantee every
infeasible interference combination will be found.

%Incompleteness is as expected: the procedure is a quick way of pruning
%obviously-infeasible combinations before the more computationally expensive
%thread-level analysis. 
%
Incompleteness is expected: the intent is a quick pruning of infeasible
combinations before the computationally expensive thread-level analysis.
%
%Insisting on completeness would be
%\emph{over-kill}: the overhead outweighs the benefit.  
%
The overhead of insisting on completeness would outweigh its benefit: the
feasibility checking problem, in the worst case, is as hard as program
verification itself, which is undecidable.
%
% After all, the feasibility checking problem, in the worst case, is as hard as
% program verification itself, which is undecidable.

Now, we formally state the soundness of our deductive procedure.
First, our deduction of the \textsc{NoReorder} relation relaxing the program
order requirement, from Figure~\ref{fig:wmm}, is sound.

\begin{theorem}
%Given two instructions $s_1$ and $s_2$ in same thread.  
Let $s_1$ and $s_2$ be two instructions in the same thread.
If our rules deduce $(s_1,s_2)\in\textsc{NoReorder}$, then the
reordering of $s_1$ and $s_2$ is not allowed by the corresponding
memory model.
\end{theorem}

The proof of this theorem is straightforward, since our inference
rules for deducing \textsc{NoReorder} directly follow the memory model
semantics provided by Adve and Gharachorloo~\cite{AdveG96} in
Figure~\ref{fig:wmm}.

Next, we note that, given the \textsc{ReadsFrom} relation, the deduction of the
\textsc{MustNotReadFrom} relation is also sound.

\begin{theorem}
Let $l$ and $s$ be two instructions.  If our rules deduce to
$(l,s)\in\textsc{MustNotReadFrom}$, then $l$ cannot read from $s$.
\end{theorem}

The proof of this theorem is straightforward: it amounts to
proofs of Rules (1)--(6).  
%
%During our presentation of these rules, we have already explained why they are
%correct.  
%
During the previous presentation, we have argued why each rule is correct.
%
%A more formal proof of their correctness can be obtained
%through \emph{proof-by-contradiction}, i.e., by first assuming a rule
%is incorrect and then deducing a contradiction.  Since this is a
%straightforward process, we omit the details for brevity.
More formal proofs can be obtained via \emph{proof-by-contradiction}, which is straightforward. We omit the details for
brevity.

\subsection{Relaxing the Write-Atomicity Requirement}
\label{sec:write-atom}

Our method soundly models \emph{buffer forwarding}, which corresponds to the
write-atomicity requirement (Column~10 Figure~\ref{fig:wmm}).  
This allows a thread to read its own write before the written value is flushed
to the memory, thus becoming visible to other threads. 
This is modeled in both the thread-level analyzer (\AnalyzeSeq) and the
deduction rules.

\AnalyzeSeq{} captures the relaxation for free. During this analysis each thread
is treated as a sequential program: all loads read their values from the
preceding writes within the same thread.

%In the deduction rules when defining \textsc{NoReorder} for TSO, PSO, and RMO
%(Section~\ref{sec:program-order}), we allowed a store to be reordered with a
%subsequent load of the same variable (Column~4 of Figure~\ref{fig:wmm}).  
%
The deduction rules for \textsc{NoReorder} (Section~\ref{sec:program-order}) always
permit the reordering of a store with a subsequent load of the same variable
(Column~4, Figure~\ref{fig:wmm}).
That is, if
$(s_1,v_1)\in\textsc{IsStore}$ and $(s_2,v_1)\in\textsc{IsLoad}$, we
do not deduce $(s_1,s_2)\in\textsc{NoReorder}$ due to buffer forwarding (even though it is counter-intuitive).
Within a thread $t$,
it may appear to be the case that the store and load are reordered from the perspective of all
threads $t' \neq t$.
Forbidding this reordering would be  equivalent to forcing a full flush of the
store-buffers before every load, thus prohibiting any thread
from reading its own store earlier than other threads.
%(i.e., before other threads observe the store). 

%%%%%%% Begin write atomicity example

\begin{figure}[htb]
\centering

{\footnotesize

\begin{minipage}{0.425\linewidth}
\begin{lstlisting}[language=C,frame=single]
void thread1() {
  x = 1; // s1
  a = x; // s2
  b = y; // s3
}
\end{lstlisting}
\end{minipage}
\hspace{0.10\linewidth}
\begin{minipage}{0.425\linewidth}
\begin{lstlisting}[language=C,frame=single]
void thread2() {
  y = 1; // s4
  fence; // s5
  c = x; // s6
}
\end{lstlisting}
\end{minipage}

\textcode{assert( !(a ==1 \&\& b == 0 \&\& c == 0) );}
}

%\caption{Example for write-atomicity requirement under TSO: we assume
%  both $x$ and $y$ are set to 0 initially.}
\caption{Write atomicity example under TSO.}
\label{fig:popl}
\vspace{-2ex}
\end{figure}

%Figure~\ref{fig:popl} shows a concrete example that explains why we need the
%relaxation to permit \emph{read-own-write} behaviors.  
%
Figure~\ref{fig:popl} exemplifies the requirement of this relaxation.
First, the assertion may be violated under TSO.  An error trace is:
\textcode{x = 1; a = 1; b = 0; y = 1; flush y; c = 0; flush x}.
To permit this trace, we must allow the following interference combination:
$s_2$ reads from $s_1$, $s_3$ reads from the initial value 0, and $s_6$ reads
from the initial value 0.  
This combination is feasible only when we avoid enforcing the program order
between $s_1$ and $s_2$.
Specifically, the statements in thread 2 follow program order ($s_4, s_5, s_6$)
from the fence. In thread 1, $s_2$ and $s_3$ are ordered since they are added
to \textsc{NoReorder} under TSO.  But, statements $s_1$ and $s_2$
are not added to \textsc{NoReorder}, thus preventing the assertion from being
(incorrectly) verified.

%%%%%%% End write atomicity example

%\mknote{In order to give full treatment to the example I've added it to the end
%  of section 4.5. We need both the treatment of fences, and the deduction of
%MHB in order to explain the reviewers counterexample fully.}

%\section{Overall Algorithm for Thread-Modular Abstract Interpretation}
\section{The Thread-Modular Analysis}
\label{sec:algorithm}

Next, we present the integration of our interference analysis
(Section~\ref{sec:feasibility}) with a thread-modular analyzer.
The thread-modular analyzer itself is standard, whose full details may
be found in several prior works including \cite{Mine12,Mine14} and
\cite{KusanoW16}.  Thus, our presentation of the analyzer itself will
be terse.
Instead, we shall focus on our main contribution, which is adding the
capability of deducing infeasible interference combinations for
weak-memory models: our method is sound for not only SC but also TSO,
PSO, and RMO. Prior techniques were either MM-oblivious, or sound only
for SC.
%Our presentation will be terse because the algorithm shares many
%structural similarities with prior thread-modular abstract
%interpreters such as Min\'{e}'s method~\cite{Mine14} and
%\Watts{}~\cite{KusanoW16}.  However, our method is sound for both SC
%and weaker memory models such as TSO, PSO, and RMO
%(Section~\ref{sec:feasibility}), whereas the prior techniques were
%either MM-oblivious or sound for SC only.

%The thread-modular introduces the notion of \emph{interferences}.
%Next, we define interferences within the thread-modular analysis.
%
Given a load $l$ of $v$, the \emph{interferences} on $l$, within the
thread-modular analysis, are the environments after all stores to $v$ from
other threads.
The function
$\GraphNodes(G)$ takes a graph as input and returns the nodes of the
graph.
%
%We collect all the memory environments from interfering stores associated with
%each load within a thread $G$ via the least fixed-point of the function
%$\Interfs'$:
The interferences on the loads in a thread $G$ is the least
fixed point of the function $\Interfs'$.
\begin{align*}
  &\begin{aligned}%
    &\Interfs'(G, M, I) = l \mapsto \{e\} \cup I(l) \\
  \end{aligned}\\
  &\begin{aligned}
  \mathbf{where\:} 
                  & e \text{ is the environment after a remote store } 
                    s \not\in\GraphNodes(G)  \\
                  & \text{ to the same variable as loaded by load } l \in
                    \GraphNodes(G) \\
    %\mathbf{where\:} & v \text{ is the variable stored into by node } 
    %                     n\not\in \GraphNodes(G) \text{, and  } \\
    %                 & e = M(n) \text{ is memory environment after $n$.}
  \end{aligned}\\
  %& \Interfs \in \GraphSet \rightarrow (\NodeSet \mapsto \StateSet) 
  %\rightarrow (\NodeSet \mapsto \PowSet(\StateSet)) \\
  &\Interfs(G, M) = \lfp(\Interfs'(G, M), I_\bot)
\end{align*}
We use $\Interfs'(G,M,I)$ as shorthand for $\Interfs'(G,M)(I)$, where
$\Interfs'(G,M)$ is a partially-applied function,  and use $I_\bot$ as the
initial map from loads to interfering-environments, i.e., one mapping all nodes
to $\{\bot\}$.
$\lfp$ computes the least fixed point.
%The function $\lfp \in (A \rightarrow A) \rightarrow A \rightarrow A$ takes a
%function $f$ and an initial value $i$ and computes the least fixed-point of $f$
%using $i$.
%
$\Interfs'$ depends on the existence of $M$, a map from each program location,
in all threads, to an environment.
We show shortly that the computation of $M$ and the interferences is done in a
nested fixed point.

We refer to an \emph{interference combination}, $\mathit{ic} \in (\NodeSet
\mapsto \StateSet$), as a map from a load $l$ to the memory environment after a
store instruction from which $l$ reads.
This differs slightly from the definition of Section~\ref{sec:feasibility} where
it is defined as a set of load-stores pairs. 
The two can be easily converted as the analysis keeps track of all the
environments associated with each store.
Given the set of interferences $I$ from $\Interfs$, the set of all
interference-combinations are all permutations of selecting
a single environment from $I$ for each load.
%
%\footnote{While this process is
%  exponential wrt to the number of loads it can increase accuracy, and be
%mitigated by optimizations and heuristics~\cite{KusanoW16}.}
%
The iterative thread-modular analysis separately considers each
interference-combination thus increasing accuracy.

The \emph{thread analyzer} adapts the sequential analyzer ($\AnalyzeSeq$,
Section~\ref{sec:background}) to use interference-combinations.
$\AnalyzeTM'$ takes a thread $G$ and an interference combination $\mathit{ic}$
and computes the input environment $e$ for some node $n$ in $G$ by joining the
environment after the predecessors of $n$ with $n$'s environment in $ic$,
denoted $ic(n)$.
Then, $e$ is passed to $n$'s transfer function to update $M(n)$.
%
%%The analyzer computes, for each thread $G$, a map $M\in(\NodeSet \mapsto
%%\StateSet)$ from every program location $n\in\NodeSet$ to its memory
%%environment. 
%%
%The function $\AnalyzeTM$ takes as input the graph of a thread $G$, an
%interference combination $\mathit{ic} \in (\NodeSet \mapsto \StateSet)$, and a
%memory map $M \in (\NodeSet \mapsto \StateSet)$. 
%%
%$M$ is updated (cf.\ $\AnalyzeSeq$) by first computing the input
%environment, $e$,  for some node $n$ in $G$ by joining the
%environment after the predecessors of $n$, and, also, the querying the interference
%combination, $\mathit{ic}$, for an interfering memory, if any.
%%
%$e$ is then passed to $n$'s transfer function to update $M(n)$.
%
%The existing procedure takes a map $M$ as input; it transform the map
%into a new map $M'$, until a fixed-point is reached ($M' = M$).  In our
%new analyzer, we add the interference combination $ic\in (\NodeSet
%\mapsto \StateSet)$ as another input.%
%
% Analyze TM with where (multiline)
\begin{align*}%
& \AnalyzeTM'(G, \mathit{ic}, M) = n \mapsto \tfunc(n)(e) \\
&\begin{aligned}
  \mathbf{where\:} & e = \bigsqcup_{(n',n) \in \GraphTrans(G)}M(n') \sqcup \mathit{ic}(n)
\end{aligned} \\
&\AnalyzeTM(G, \mathit{ic}) = \lfp(\AnalyzeTM'(G, \mathit{ic}), M_\bot)%
\end{align*}%
%
% Single line AnalyzeTM (breaks the margins)
%\begin{equation*}%
%\AnalyzeTM'(G, \mathit{ic}, M) = n \mapsto \tfunc(n)(\bigsqcup_{(n',n) \in \delta}M(n') \sqcup \mathit{ic}(n)) \\
%\end{equation*}%
%
$\AnalyzeTM$ is the least fixed point of $\AnalyzeTM'$.
$\GraphTrans(G)$ returns the transition-relation of a graph.
$M_\bot$ is the initial memory map mapping the entry nodes of each thread to
$\top$ and all others to $\bot$.
Given a set of threads $\mathit{Gs}$ and a set of interference-combinations
$I$, applying $\AnalyzeTM$ to each $G \in \mathit{Gs}$ and each
$\mathit{ic} \in I$ computes the analysis over all threads.

What remains is to show how the thread analyzer and the calculation of
interferences can be done simultaneously since they are dependent: the
interference computation depends on the analysis result, $M$, and the analysis
result depends on the set of interferences, $I$.
The solution is a nested fixed point: the outer computation produces $M$, and
the inner computation produces $I$. The process iterates until $M$ (and thus
$I$) reach a fixed point.
\begin{align*}%
  &\Analyze(G, M) = M' \\
  &\begin{aligned}
    %\mathbf{where\:} & I \in \NodeSet \rightarrow \PowSet(\StateSet) \\
    \mathbf{where\:} & I = \Interfs(G, M) \\
                     %& \mathit{I'} \in \PowSet(\NodeSet \rightarrow \PowSet(\StateSet)) \\
                     & \mathit{I'} = \FilterFeasible(I) \\
                     % Cs :: P(N -> P(S))
                     %& \mathit{Cs} \in \PowSet(\NodeSet \rightarrow \PowSet(\StateSet)) \\
                     %& \textcolor{blue}{\mathit{Cs} = \CartProd( \LoadToInterfs(G, I)  )} \\
                     %& \mathit{Cs'} \in \PowSet(\NodeSet \rightarrow \PowSet(\StateSet)) \\
                     %& \textcolor{blue}{\mathit{Cs}' = \FilterFeasible(\mathit{Cs})     } \\
                     %& M' \in (\NodeSet \mapsto \StateSet) \\
                     & M' = \JoinMM(\map(\AnalyzeTM(G), I')) \\
  \end{aligned} \\
  %&\Analyze(G, M) = \lfp(\Analyze'(G), M) \\
  &\AnalyzeAll'(\mathit{Gs}, M)  = \JoinMM(\map(\Analyze(M), \mathit{Gs}))\\
  &\AnalyzeAll(\mathit{Gs})  = \lfp(\AnalyzeAll'(\mathit{Gs}), M_\bot)
\end{align*}
$\Analyze$ operates as follows: first, it takes $M$, the current analysis results over
all threads, and computes the interferences, $I$, wrt the thread under test, $G$.
The function $\FilterFeasible$ integrates the thread-level analyzer with the
feasibility analysis of Section~\ref{sec:feasibility}. It expands the
interferences $I$ into a set of interference combinations $I'$, and filters
any infeasible combination. 

Specifically, given the interferences on a thread, $I = \{(l_1 \mapsto \{e_1,
e_2, \ldots\}), (l_2 \mapsto \{e_3, e_4, \ldots\})\ldots\}$, $\FilterFeasible$ 
creates all combinations of pairing each load to a single interfering
environment, e.g., 
$I_e = \{\{\langle l_1, e_1 \rangle, \langle l_2, e_3 \rangle, \ldots\}
,\{\langle l_1, e_2 \rangle, \langle l_2,e_3 \rangle, \ldots\}
,\ldots$\}.
Then, it maps each environment in $I_e$ to the associated store generating the
environment, e.g., $I_s = \{\{\langle l_1, s_1 \rangle, \langle l_2, s_3
\rangle\}, \ldots\}$.
Each set of pairs of load and store statements in $I_s$ is then passed to the
deduction analysis of Section~\ref{sec:feasibility}. If it is infeasible, it is
discarded, otherwise it is added to the set $I'$ returned by $\FilterFeasible$.

%
%Note that while $\Analyze$ is from Kusano and Wang~\cite{KusanoW16}, their
%filtering procedure was unsound for programs written under memory-models weaker
%than SC.
%
$\map(\AnalyzeTM(G), I') \in 2^{(\NodeSet \mapsto \StateSet)}$ is the set
of the results of applying $\AnalyzeTM(G, i)$ for each $i \in I'$.
Specifically, $\map \in (A \rightarrow B) \rightarrow 2^A
\rightarrow 2^B$ takes a function $f$ and a set $S$,
and returns a set containing the application of $f$ on each element of $S$.

$\JoinMM \in (2^{(\NodeSet \mapsto \StateSet)} \rightarrow (\NodeSet \mapsto
\StateSet))$ takes the join of memory environments on matching nodes across a
set of maps to join them into a single map.
Similarly, $\AnalyzeAll'$ joins the results of applying $\Analyze$ to the set
$\mathit{Gs}$ of threads. $\AnalyzeAll$ computes the fixed point of $\AnalyzeAll'$ starting
with the initial map $M_\bot$.
%
% Analyzing a set of threads $\mathit{Gs}$ is $\lfp (\AnalyzeAll'(\mathit{Gs}))$
% where $\AnalyzeAll \in (\NodeSet \rightarrow \StateSet) \rightarrow (\NodeSet
% \rightarrow \StateSet)$. $\AnalyzeAll'$ joins the result of calling  $\Analyze$
% on each individual thread.

The following is a high-level example. Initially, each thread $G$ is analyzed
in the presence of $M_\bot$ resulting in the set of interferences, $I$, being
empty (all stores map to $\bot$). The results of analyzing each thread are
merged into a new map $M$. Each thread is then analyzed using $M$, resulting in
the sets $I$ and $I'$ to be (potentially) non-empty, causing $\AnalyzeTM$ to
be called once per-combination. Within a thread, the results of $\AnalyzeTM$
are joined, then, across threads, the results of $\Analyze$ are joined,
creating $M'$. The procedure repeats, thus growing the size of $M$,
$I$, and $I'$ until $M = M'$.

We handle loops the same way as in prior techniques
(e.g.,~\cite{KusanoW16}).  Given a load $l$ within a loop the
previously described analysis can generate an infinite number of
interference combinations for $l$, e.g., when $l$ is within an
infinite loop.  Loops are unrolled when possible, and, when not, we
join all the \emph{feasible} interfering memory environments into a
single value.
An interfering environment $e$ is infeasible to interfere on $l$ if the store
generating $e$ must-happen after $l$; otherwise, it is feasible.
This is sound for verifying assertions embedded in a concurrent
program~\cite{KusanoW16}.

%%%%%
%Chao:  irrelevant details that no reviewer can quickly digest
%%%%%
%, but not for the
%more general case of generating sound invariants at every program
%location. To generate invariants at some statement $s$, $s$ needs to
%be considered as a load of all shared variables such that their
%potential values flow to $s$. Soundness is more fully discussed in
%Kusano and Wang~\cite{KusanoW16}.

%\newpage
\section{Experiments}
\label{sec:experiment}

We implemented our weak-memory-aware abstract interpreter
in a tool named \toolname{}, building upon
open-source platforms such as LLVM~\cite{AdveLBSG03},
Apron~\cite{Jeannet09}, and $\mu{Z}$~\cite{Hoder11}.  
Specifically, we use LLVM to translate C programs into LLVM bit-code, based on
which we perform static analysis.  We use the Apron library to manipulate
abstract domains in the thread analyzer.  We use the $\mu${Z} fixed-point
engine in Z3~\cite{DeMoura08} to solve Datalog constraints that encode the
feasibility of interference combinations.

We implemented the state-of-the-art MM-oblivious abstract
interpretation method of Min\'e~\cite{Mine14}, and the SC-specific
method, \Watts{}~\cite{KusanoW16}, on the same platform to facilitate experimental evaluation.
%
%
% %%%%
% Chao:  Irrelevant details -- only Zak (pldi reviewer) would be able to know what this means -- but he already knew.
%        Nobody else would be able to make sense of it -- but people may be puzzled by it.
% %%%%
%The analysis of Min\'e~\cite{Mine14} does not include the monotonicity domain
%or relational lock invariants.
%
%We also ran a previously implemented version of
%\Duet{}~\cite{FarzanK12,FarzanK13} (by its authors) on the same benchmarks to
%compare against the other methods.
%
We also compared against a previously implemented version of
\Duet{}~\cite{FarzanK12,FarzanK13}.
While \Duet{} may be unsound, and \Watts{} is unsound, we include their results
because they are closely related to our new technique.

All methods implemented in \toolname{} use the clustering and
property-directed optimizations~\cite{KusanoW16}, where clustering
considers interferences only within sets of loads, similar to the
packing of relational domains, and property-direction filters
interference combinations unrelated to properties under
test. These optimizations reduce the number of interference
combinations, which is crucial since it grows exponentially
with respect to program's size.

We evaluated \toolname{} on a large set of programs written using the
POSIX threads.  These benchmarks fall into two categories.  The first are 209
litmus tests exposing non-SC behaviors under various processor-level memory
models~\cite{Alglave13T}.  
The second are 52 larger applications~\cite{svcomp15,LinuxISR,FarzanK12},
including several Linux device drivers. 
%
%In total, these benchmark programs have 61,981 lines of code.
The benchmarks total 61,981 lines of code.
The properties under verification are assertions embedded in the
program's source code: a property is valid if and only if the assertion holds over all
executions under a given memory model.

Our experiments were designed to answer the following research questions:
%\begin{enumerate}
%\item 
(1)
Is our new method more effective than prior techniques in obtaining
correctness proofs on relaxed memory?
%\item
(2)
Is our new method more accurate than prior techniques in detecting
potential violations on relaxed memory?
%\item 
(3)
Is our new method reasonably efficient when used as a static program
analysis technique?
%\end{enumerate}
We conducted all experiments on a Linux computer with 8 GB
RAM, and a 2.60 GHz CPU.

\subsection{Litmus Test Results}

First, we present the litmus test results.  
Since these programs are small in terms of code size, all methods under evaluation (Min\'e, \Watts{},
\Duet{}, and \toolname{}) finished quickly.  
Thus, our focus is not on comparing the runtime performance but comparing the
accuracy of their results.  
%
%We want to know if our
%method, compared to these state-of-the-art techniques, can
%significantly reduce the number of bogus alarms and bogus proofs.
%
Specifically, we compare our method to these state-of-the-art techniques in terms of the
number of true proofs, bogus proofs, true alarms, and bogus alarms.

%Here, a bogus alarm means the property holds and yet it cannot be proved.  
%
Here, a bogus alarm is a valid property which cannot be proved.
%
%Similarly, a bogus proof means the property does
%not hold and yet it is (incorrectly) proved.  Since we know \emph{a
%  priori} if a property holds or not in each of these litmus test
%programs, we have the ground truth to decide if a proof or an alarm
%reported by a method is a bogus proof or a bogus alarm.
%
A bogus proof is a property which may be violated yet is unsoundly and
incorrectly proved.
The litmus tests are particularly useful not only because they cover corner cases, but also because we know a priori if a property
holds or not.

\begin{table}[htb]
  \centering
\caption{Results on the litmus test programs under TSO.}
  \label{tbl:litmus-tso}
  \resizebox{\linewidth}{!}{%
    \begin{tabular}{lccccr}
  \toprule
  Method                  & True Alarm & Bogus Alarm &  True Proof & Bogus Proof & Time (s) \\
  \midrule
Min\'{e}~\cite{Mine14}      & 77 & 207 & 8   & 0  & 12.9\\
\Duet{}~\cite{FarzanK12}    & 77 & 181 & 34*  & 0  & 473.1 \\
\Watts{}~\cite{KusanoW16}   & 63 & 13  & %202 & 14\US{} & 71.0\\
                                         0    & 216\US{} & 71.0\\
\toolname{}                 & 77 & 72  & 143 & 0  & 89.2\\
\bottomrule
\end{tabular}
}%resizebox
\vspace{-1ex}
\end{table}

%Table~\ref{tbl:litmus-tso} shows the results of comparing all four methods
%under TSO.  
%
Table~\ref{tbl:litmus-tso} summarizes the litmus test results under TSO.
The first column shows the name of each method, and the next four show the number
of true alarms, bogus alarms, true proofs, and bogus proofs generated by each
method, respectively.
Since \Watts~\cite{KusanoW16} was designed to be SC-specific, it ignores non-SC
behaviors, meaning its proofs are unsound under weaker memory (marked by \US{}).
The last column is the total analysis time over all tests.

Overall, the results show the prior thread-modular
technique of Min\'e admits many infeasible executions thus leading to
207 bogus alarms.
\Duet{} reported 181 bogus alarms.  In contrast, our method
(\toolname{}) reported only 72 bogus alarms, together with 143 true
proofs.  Therefore, it is more accurate than these prior techniques.

Although \Watts{} reported only 13 bogus alarms, it is unsound for
TSO: it only considers SC behaviors and cannot be trusted. 
Furthermore, the soundness of \Duet{} under TSO or any other non-SC memory model
was not clear (since \Duet{} was only designed for SC). Thus, in the result table, its 34 proofs are marked with *.
%, although in these litmus
%tests, it never reported bogus proofs.

\begin{table}[htb]
  \centering
\caption{Results on the litmus test programs under PSO.}
  \label{tbl:litmus-pso}
  \resizebox{\linewidth}{!}{%
    \begin{tabular}{lccccr}
  \toprule
  Method                  & True Alarm & Bogus Alarm &  True Proof & Bogus Proof & Time (s) \\
  \midrule
Min\'{e}~\cite{Mine14}     & 81 & 203 & 8   & 0  & 12.9\\
\Duet{}~\cite{FarzanK12}   & 81 & 177 & 34*  & 0  & 473.1 \\
%\Watts{}~\cite{KusanoW16}  & 64 & 12\US  & 199 & 17\US{} & 71.0 \\
\Watts{}~\cite{KusanoW16}  & 64 & 12  & 0   & 216\US{} & 71.0 \\
\toolname{}                & 81 & 68  & 143 & 0  & 281.4 \\
   \bottomrule
\end{tabular}%
}%resizebox
\vspace{-1ex}
\end{table}

Table~\ref{tbl:litmus-pso} summarizes the results under PSO.
Again, \Watts{} may be unsound for weak memory. 
%
%Furthermore, for the first three methods, although their results under PSO
%remain the same as the under TSO, their interpretations have changed (i.e.,
%whether a reported alarm is true or bogus has changed). 
%
The same litmus programs were used under PSO as in TSO but the properties
changed, i.e., whether an alarm is true or bogus.
Note that Min\'{e} only verified 8 properties, \Duet{} verified
34, whereas our method verified 143.

\begin{table}[htb]
  \centering
\caption{Results on the litmus test programs under RMO.}
  \label{tbl:litmus-rmo}
  \resizebox{\linewidth}{!}{%
    \begin{tabular}{lccccr}
  \toprule
  Method                  & True Alarm & Bogus Alarm &  True Proof & Bogus Proof & Time (s) \\
  \midrule
Min\'{e}~\cite{Mine14}    & 28 & 67 & 8  & 0  & 4.9\\
\Duet{}~\cite{FarzanK12}  & 11 & 58 & 34 & 0  & 187.8 \\
\Watts{}~\cite{KusanoW16} & 0  & 0\US  & 75 & 28\US & 33.9\\
\toolname{}               & 28 & 13 & 62 & 0  & 46.9\\
  \bottomrule
\end{tabular}%
}%resizebox
\vspace{-1ex}
\end{table}

% Table~\ref{tbl:litmus-rmo} shows the results of comparing the four
% methods under RMO.  
%
Table~\ref{tbl:litmus-rmo} summarizes the results under RMO.
Under RMO, a different set of litmus programs were used since the instruction
set for processors using RMO differs from TSO and PSO.
Nevertheless, we observed similar results: \toolname{} obtained significantly
more true proofs and fewer bogus alarms than the other methods.

%In general, our new method was more accurate than prior techniques.
%However, like any other abstract interpretation method, it does not
%eliminate all bogus alarms due to the over-approximated nature of its
%analysis.
%
In general, our method was more accurate than prior techniques. However,
since the analysis is over-approximated, it does not eliminate all bogus
alarms.
Currently, most bogus alarms reported by \toolname{} require
reasoning across more than two threads, e.g., the correctness
of a property may require reasoning that thread $T_1$ reading $x
= 1$ from thread $T_2$ implies $y = 1$ in thread $T_3$.  
%
%Since our method is thread-modular---it analyzes one thread at a time while
%abstracting all other threads into a set of thread interferences---it cannot
%capture ordering constraints that involve more than two threads.  
%
Since our method is thread-modular---threads are analyzed individually by
abstracting all other threads into a set of interferences---it cannot capture
ordering constraints involving more than two threads.  
%
%In principle, this limitation can be lifted, e.g., by extending our deduction
%rules for checking the feasibility of interference combinations: we leave this
%as future work.
%
In principle, this limitation can be lifted by extending our 
interference feasibility analysis:  we leave this as future work.

\subsection{Results on Larger Applications}

Next, we present our results on the larger benchmark programs.  Since
execution time is no longer negligible, we compare, across methods, both the
run time and accuracy.
However, since the programs are larger (60K lines of code) and there are far
too many properties to manually inspecting each case, we do not report the number of bogus alarms and bogus
proofs due to lack of the ground truth.  
Instead, we compare the \emph{total} number of proofs reported by each method, to show our method is more accurate even though all methods are approximate.

\begin{table}
  \centering
\caption{Results on larger applications (total of 61,981 LOC).}
\label{tbl:res-tso}
\resizebox{\linewidth}{!}{%
\begin{tabular}{l*{8}{c}}
  \toprule
  & \multicolumn{2}{c}{Min\`{e}~\cite{Mine14}} 
  & \multicolumn{2}{c}{\Duet{}~\cite{FarzanK12}} 
  & \multicolumn{2}{c}{\Watts{}~\cite{KusanoW16}} 
  & \multicolumn{2}{c}{\toolname{}}\\
  \cmidrule(lr){2-3}
  \cmidrule(lr){4-5}
  \cmidrule(lr){6-7}
  \cmidrule(lr){8-9}
  Name
  & Time & Verif
  & Time & Verif
  & Time & Verif
  & Time & Verif \\
  \midrule
thread00       & 0.01 & 0  & 1.1 & 0  &  0.03 & 0   & 0.03 & 0  \\%& 0.04 & 0 & 0.03 & 0 \\
threadcreate01 & 0.02 & 1  & 0.8 & 1  &  0.05 & 2   & 0.04 & 2  \\%& 0.04 & 2 & 0.04 & 2 \\
threadcreate02 & 0.02 & 1  & 0.7 & 2  &  0.03 & 2   & 0.03 & 2  \\%& 0.03 & 2 & 0.03 & 2 \\
sync\_01\_true & 0.03 & 1  & 1.3 & 1  &  0.06 & 1   & 0.06 & 1  \\%& 0.06 & 1 & 0.06 & 1 \\
sync\_02\_true & 0.03 & 1  & 1.2 & 1  &  0.07 & 1   & 0.07 & 1  \\%& 0.07 & 1 & 0.07 & 1 \\
intra01        & 0.02 & 1  & 1.2 & 0  &  0.04 & 2   & 0.04 & 2  \\%& 0.04 & 2 & 0.04 & 2 \\
dekker1        & 0.10 & 3  & 1.3 & 2  &   6.1 & 4   &  6.6 & 4  \\%& 6.52 & 4 & 6.53 & 4 \\
fk2012\_v2     & 0.04 & 3  & 1.4 & 3  &   0.3 & 4   &  0.2 & 4  \\%& 0.23 & 4 & 0.22 & 4 \\
keybISR        & 0.04 & 4  & 1.2 & 4  &   2.1 & 6   &  2.3 & 5  \\%& 2.27 & 5 & 2.26 & 5 \\
ib700wdt\_01   &  1.7 & 23 & 1.8 & 35 &  14.5 & 46  & 11.4 & 46 \\%& 11.51 & 46 & 11.41 & 46 \\
ib700wdt\_02   & 13.2 & 63 & 2.7 & 95 & 108.4 & 126 & 89.4 & 126 \\%& 89.29 & 126 & 89.96 & 126 \\
ib700wdt\_03   & 23.1 & 81 & 2.8 & 122& 178.4 & 162 &156.3 & 162 \\%& 155.79 & 162 & 154.60 & 162 \\
i8xx\_tco\_01  &  1.0 & 14 & 2.8 & 28 &  97.4 & 39  & 53.3 & 39 \\%& 54.8 & 39 & 53.03 & 39 \\
i8xx\_tco\_02  &  8.5 & 34 & 4.6 & 68 &1288.3 & 99  &757.3 & 99 \\%& 746.1 & 99 & 751.56 & 99 \\
i8xx\_tco\_03  & 10.7 & 37 & 4.7 & 74 &1677.0 & 108 &952.3 & 108 \\%& 962.9 & 108 & 956.54 & 108 \\
machzwd\_01    &  0.6 & 14 & 1.3 & 14 & 107.5 & 35  & 42.1 & 34 \\%& 43.5 & 34 & 41.29 & 34 \\
machzwd\_02    &  1.6 & 24 & 1.4 & 24 & 240.5 & 65  & 75.3 & 64 \\%& 73.49 & 64 & 74.87 & 64 \\
machzwd\_03    &  4.1 & 39 & 1.8 & 39 & 488.7 & 110 &128.9 & 109 \\%& 125.33 & 109 & 125.41 & 109 \\
mixcomwd\_01   &  0.8 & 12 & 1.9 & 21 & 169.3 & 34  & 15.3 & 33 \\%& 15.58 & 33 & 15.65 & 33 \\
mixcomwd\_02   &  2.2 & 32 & 4.4 & 41 & 768.0 & 64  & 61.7 & 63 \\%& 65.86 & 63 & 63.12 & 63 \\
mixcomwd\_03   &  3.5 & 64 & 2.9 & 65 &  88.3 & 100 & 84.0 & 100 \\%& 84.09 & 100 & 84.02 & 100 \\
pcwd\_01       &  0.7 & 10 & 0.9 & 22 &   1.8 & 31  &  1.4 & 31 \\%& 1.42 & 31 & 1.43 & 31 \\
pcwd\_02       &  4.4 & 27 & 1.3 & 56 &   8.5 & 82  &  6.8 & 82 \\%& 6.84 & 82 & 6.79 & 82 \\
pcwd\_03       & 10.2 & 40 & 1.7 & 82 &  18.5 & 121 & 15.4 & 121 \\%& 15.37 & 121 & 15.39 & 121 \\
pcwd\_04       & 25.4 & 60 & 2.1 & 122&  46.1 & 181 & 39.8 & 181 \\%& 39.68 & 181 & 39.87 & 181 \\
sbc60xxwdt\_01 &  1.1 & 21 & 2.0 & 0  &   4.9 & 43  &  3.0 & 43 \\%& 3.13 & 43 & 3.13 & 43 \\
sbc60xxwdt\_02 &  3.3 & 40 & 2.3 & 0  &  13.7 & 81  &  8.0 & 81 \\%& 8.08 & 81 & 8.00 & 81 \\
sbc60xxwdt\_03 &  7.0 & 60 & 3.2 & 0  &  33.4 & 121 & 17.9 & 121 \\%& 17.71 & 121 & 17.86 & 121 \\
sc1200wdt\_01  &  1.0 & 22 & 1.3 & 10 &  15.1 & 33  & 10.6 & 33 \\%& 10.62 & 33 & 10.69 & 33 \\
sc1200wdt\_02  &  6.9 & 58 & 2.2 & 28 &  64.5 & 87  & 47.6 & 87 \\%& 47.97 & 87 & 47.72 & 87 \\
sc1200wdt\_03  & 26.2 & 102& 3.3 & 50 & 197.0 & 153 &146.7 & 153 \\%& 146.85 & 153 & 147.47 & 153 \\
smsc37b787wdt\_01&1.0 & 22 & 1.2 & 23 &  16.7 & 46  & 12.0 & 46 \\%& 12.21 & 46 & 12.15 & 46 \\
smsc37b787wdt\_02&9.2 & 76 & 2.3 & 77 &  91.4 & 154 & 67.3 & 154 \\%& 67.84 & 154 & 66.83 & 154 \\
smsc37b787wdt\_03&26.4& 130& 3.4 & 131& 286.1 & 262 &197.5 & 262 \\%& 196.97 & 262 & 197.57 & 262 \\
sc520wdt\_01   &  1.5 & 15 & 1.1 & 16 &  15.6 & 45  & 11.0 & 45 \\%& 11.06 & 45 & 11.08 & 45 \\
sc520wdt\_02   & 12.6 & 41 & 1.7 & 42 &  74.1 & 123 & 56.0 & 123 \\%& 56.13 & 123 & 55.12 & 123 \\
sc520wdt\_03   & 27.8 & 58 & 2.2 & 59 & 155.6 & 174 &116.9 & 174 \\%& 119.43 & 174 & 127.72 & 174 \\
w83877fwdt\_01 & 12.5 & 34 & 1.8 & 34 &  83.9 & 137 & 71.0 & 137 \\%& 69.43 & 137 & 68.95 & 137 \\
w83877fwdt\_02 & 29.0 & 50 & 2.3 & 50 & 189.6 & 201 &159.6 & 201 \\%& 159.72 & 201 & 161.76 & 201 \\
w83877fwdt\_03 & 54.2 & 66 & 2.7 & 66 & 357.9 & 265 &301.9 & 265 \\%& 294.15 & 265 & 292.12 & 265 \\
wdt01          &  0.2 & 3  & 1.4 & 13 &  65.9 & 14  & 27.6 & 14 \\%& 27.44 & 14 & 27.14 & 14 \\
wdt02          &  0.3 & 5  & 1.5 & 21 & 600.2 & 22  &306.5 & 22 \\%& 292.21 & 22 & 319.17 & 22 \\
wdt03          &  0.5 & 6  & 1.5 & 25 &1479.1 & 26  &766.2 & 26 \\%& 785.61 & 26 & 739.80 & 26 \\
wdt977\_01     &  1.3 & 9  & 1.9 & 35 &  83.0 & 43  & 49.7 & 43 \\%& 50.70 & 43 & 50.05 & 43 \\
wdt977\_02     &  2.4 & 13 & 2.3 & 51 & 115.6 & 63  & 76.0 & 63 \\%& 76.07 & 63 & 76.22 & 63 \\
wdt977\_03     &  8.2 & 25 & 2.9 & 99 & 264.9 & 123 &193.9 & 123 \\%& 193.83 & 123 & 193.72 & 123 \\
wdt\_pci01     &  1.2 & 11 & 1.1 & 31 &   5.9 & 52  &  4.6 & 52 \\%& 4.75 & 52 & 4.66 & 52 \\
wdt\_pci02     &  8.9 & 31 & 1.8 & 91 &  33.2 & 152 & 26.3 & 152 \\%& 26.23 & 152 & 26.01 & 152 \\
wdt\_pci03     & 23.9 & 51 & 3.0 & 151&  93.5 & 252 & 72.7 & 252 \\%& 72.17 & 252 & 71.68 & 252 \\
pcwd\_pci\_01  &  4.7 & 56 & 1.3 & 89 &  27.2 & 116 & 21.1 & 116 \\%& 21.15 & 116 & 21.19 & 116 \\
pcwd\_pci\_02  &  9.8 & 70 & 1.4 & 132&  52.6 & 158 & 40.6 & 158 \\%& 40.74 & 158 & 40.88 & 158 \\
pcwd\_pci\_03  & 20.3 & 88 & 2.0 & 186&  97.7 & 212 & 72.7 & 212 \\%& 72.37 & 212 & 72.56 & 212 \\
  \midrule
Total          & 415 s & {\bf 1752} & 106 s & {\bf 2432}*   & 9830 s & {\bf 4583}\US   & 5387 s & {\bf 4577}   \\%& 5385.58 & 4577   & 5365.47 & 4577 \\
  \bottomrule

\end{tabular}
}%resizebox
\vspace{-1ex}
\end{table}

Table~\ref{tbl:res-tso} shows our results under TSO, where \US{} and *
mark the unsoundly verified properties.  Since the results for PSO and
RMO are similar to Table~\ref{tbl:res-tso}, we omit them for brevity.
Column~1 of this table shows the name of the benchmark program.  
%
%Columns~2--3 show the time taken by Min\'e's method (in seconds) and the number
%of properties it was able to prove.  
%%
%Columns~4--5 show the time taken by \Duet{} and the
%number of properties it was able to prove.  
%
%Columns~6--7 show the time taken by \Watts{} and the number of proved
%properties.  
%Finally, Columns~8--9 show the time taken by \toolname{} and the number of
%proved properties.
%
Columns 2--3, 4--5, 6--7, and 8--9 show the run time and number of properties verified
by Min\'e, \Duet, \Watts, and \toolname{}, respectively.

Again, while the proofs reported by \toolname{} and Min\'e's
method are sound, the proofs reported by \Watts{} are
not, and the soundness of \Duet{} on weak memory is unclear.

%Overall, our method (\toolname{} was able to prove 4,577 properties,
%compared to 4,583\US properties proved by \Watts{} (including bogus
%proofs), 2,432 properties proved by \Duet{}, and only 1,712 properties
%proved by Min\'{e}'s method.
%
Overall, \toolname{} proved 4,577 properties compared to only 1,712 
proved by Min\'e, an increase of 2.7x more properties relative to prior
state-of-the-art.
Additionally, though \Duet{} may be unsound, it proved only 2,432 properties.
The definitely-unsound \Watts{} ``proved'' 4,583 properties, possibly including bogus
proofs.
%
%Although the number of proofs obtained by \Watts{} is close to ours,
%they cannot be trusted due to the unsoundness of its modeling of
%non-SC behaviors.

In terms of the run time, \toolname{} took 5,387 seconds, which is
similar to \Watts{}, and slower
than \Duet{} and Min\'{e}.
However, the additional time is well justified due to the significant increase
in the number of proofs.  
Furthermore, the runtime performance -- proving 1 property per second -- remains
competitive as a static analysis technique.

To summarize, our new method has modest runtime overhead compared to
prior techniques, but vastly improved accuracy in terms of the
analysis results, and is provably sound in handling not only SC but
also three other processor-level memory models.

\section{Related Work}
\label{sec:related}

We reviewed prior work on thread-modular abstract interpretation,
which are either MM-oblivious~\cite{Mine11,Mine12,Mine14} or
SC-specific~\cite{KusanoW16} in processor memory-models.  There are
also techniques~\cite{FarzanK12,FarzanK13,RoychoudhuryM02,HuynhR06}
that are not thread-modular.

There are code-transformation
techniques~\cite{KupersteinVY11,Meshman2014,Dan15} that transform a
non-SC program into an SC program and then apply abstract
interpretation.
They generally follow the \emph{sequentialization} approach pioneered by Lal
and Reps~\cite{LalR09}, with a focus on code transformation as opposed to
abstract interpretation. 
To ensure termination, they make various assumptions to bound the
program's behavior. Furthermore, they are not thread-modular, and
often do not directly handle C code.
Instead, they admit only \emph{models} of concurrent programs written in
artificial languages; because of this, we were not able to perform a direct
experimental comparison.

In the context of bounded model checking, Alglave et al.\ proposed
several methods for concurrent software on relaxed memory.
They are based on either sequentializing concurrent
programs~\cite{Alglave13T} or encoding weak memory semantics 
using SAT/SMT solvers~\cite{Alglave13P,Alglave14}.  Alglave et
al.\ also developed techniques for modeling and testing weak-memory
semantics of real processors~\cite{AlglaveMT14}, and characterized the
memory models of some GPUs~\cite{AlglaveBDGKPSW15}.  However, these
techniques are primarily for detecting buggy behaviors as opposed to
proving that such behaviors do not exist.

In the context of systematic testing, often based on stateless model
checking~\cite{Godefr97,Flanag05,WangSG11} or predictive
analysis~\cite{WangKGG09,SaidWYS11,SinhaMWG11,SinhaMWG11hvc,WangG11,HuangMR14}, a number of methods have been proposed to
handle weak memory such as
TSO/PSO~\cite{ZhangKW15,AbdullaAAJLS15,DemskyL15,Huang016},
PowerPC~\cite{AbdullaAJL16}, and C++11~\cite{NorrisD13,OuD17}.  However,
since they rely on concretely executing the program, and
require the user to provide test inputs, they can only be
used to detect bugs.  That is, since testing does not cover all
program behaviors, if no bug is detected, these methods cannot obtain
a correctness proof.  In contrast, our method is based on abstract
interpretation, which covers all possible program behaviors and
therefore is geared toward obtaining correctness proofs.

Thread-modular analysis was also used in model
checking~\cite{FlanaganFQ02,FlanaganQ03}, where it was combined with
predicate abstraction~\cite{HenzingerJMQ03} to help mitigate state
explosion and thus increase the scalability.  However, model checking
is significantly different from abstract interpretation in that each
thread must be first abstracted into a finite-state model.
Thread-modular analysis was also used to conduct shape
analysis~\cite{GotsmanBCS07} and prove thread
termination~\cite{CookPR07}. Hoenicke et al.~\cite{HoenickeMP17}
introduced a hierarchy of proof systems that leverage thread
modularity in compositional verification on SC memory.

Similar to the interference analysis in Watts~\cite{KusanoW16}, we check
the feasibility of thread interactions using Datalog.  Datalog-based
declarative program analysis was a framework introduced by Whaley and
Lam~\cite{WhaleyL04}. Previously, it has been used to implement
points-to~\cite{LamWLMACU05,BravenboerS09},
dependency~\cite{GuoKWYG15,SungKSW16} and change-impact
analyses~\cite{GuoKW16}, uncover security bugs~\cite{LivshitsL05} and
detect data races~\cite{NaikAW06}.

In abstract interpretation of sequential programs, Min\'{e}~\cite{Mine06} proposed a technique
for abstracting the global memory into a set of byte-level cells to
support a variety of casts and union types.
Ferrara et al.~\cite{Ferrara14,Ferrara15} integrated heap abstraction
and numerical abstraction during static analysis, where the heap is
represented as disjunctions of points-to constraints based on values.
Jeannet and Serwe~\cite{Jeannet04} also proposed a method for
abstracting the data and control portions of a call-stack for
analyzing sequential programs with potentially infinite recursion.
Subsequently, Jeannet~\cite{Jeannet12} extended the work to handle concurrent
programs as well.
However, none of these methods was designed specifically for
handling weak memory models.

\section{Conclusions}
\label{sec:conclusion}

We have presented a thread-modular static analysis method for concurrent
programs under \emph{weak memory models}, building upon a lightweight
constraint system for quickly identifying the infeasibility of thread
interference combinations, so they are skipped during the expensive
abstract-interpretation based analysis.  The constraint system
is also general enough to handle a range of processor-level memory
models.  We have implemented the method and conducted experiments on a
large number of benchmark programs.  We showed the new method
significantly outperformed three state-of-the-art techniques in terms
of accuracy while maintaining only a moderate runtime overhead.

\section*{Acknowledgments}

This material is based upon research supported in part by the
U.S.\ National Science Foundation under grants CNS-1405697 and
CCF-1722710 and the U.S.\ Office of Naval Research under award number
N00014-13-1-0527.

\clearpage
\newpage
\bibliographystyle{plain}

\begin{flushleft}
\bibliography{mk}
\end{flushleft}

\end{document}